\documentclass[usenatbib]{mn2e}
\usepackage{amsmath}
\usepackage{txfonts}
\usepackage[pdftex]{graphicx}
\usepackage{array}

\newcommand{\oiii}{[O\,{\sc iii}]}

\newcommand{\civ}{C\,{\sc iv}}

\newcommand{\ci}{C\,{\sc i}}

\newcommand{\mgii}{Mg\,{\sc ii}}
\newcommand{\co}{CO\,{\sc (1$\rightarrow$0) }}

\newcommand{\hst}{{\sl HST} }
\newcommand{\vla}{{\sl VLA} }
\newcommand{\evla}{{\sl JVLA } }

\newcommand{\vlba}{{\sl VLBA } }
\newcommand{\merlin}{{\sl MERLIN } }
\newcommand{\gmrt}{{\sl GMRT } }
\newcommand{\ami}{{\sl AMI } }

\newcommand{\mnras}{MNRAS}
\newcommand{\nat}{Nat}
\newcommand{\aj}{AJ}
\newcommand{\apj}{ApJ}
\newcommand{\apjl}{ApJL}
\newcommand{\aap}{A\&A}

\newcommand{\araa}{ARA\&A}
\newcommand{\pasj}{PASJ}

\newcommand{\na}{New Astronomy}
\newcommand{\bs}{\boldsymbol}

\title[Preferential magnification in IRAS~FSC10214+4724  \--- I.]{\mbox{The preferentially magnified active nucleus in IRAS\,F10214+4724\,\---\,I.} Lens model and spatially resolved radio emission}

\author[Deane et al.]{R.P. Deane$^{1}$\thanks{E-mail: roger.deane@astro.ox.ac.uk}, S. Rawlings$^{1}$, P.J. Marshall$^{1}$, I. Heywood$^{1}$, H.-R. Kl\"ockner$^{1,2}$, 
\newauthor K. Grainge$^{3,4}$, T. Mauch$^{1,5}$, S. Serjeant$^{6}$
\vspace*{3pt}\\
\noindent $^1$Astrophysics, Department of Physics, University of Oxford, Keble Road, Oxford, OX1 3RH, UK \\
\noindent $^2$Max-Planck-Institut f\"ur Radioastronomie, Auf dem H\"ugel 69, 53121 Bonn, Germany \\ 
\noindent $^3$Astrophysics Group, Cavendish Laboratory, J J Thomson Avenue, Cambridge CB3 0HE, UK\\
\noindent $^4$Kavli Institute for Cosmology Cambridge, Madingley Road, Cambridge, CB3 0HA, UK \\
\noindent $^5$Centre of Astrophysics Research, University of Hertfordshire, Hatfield, AL10 9AB, UK \\
\noindent $^6$Department of Physics and Astronomy, The Open University, Milton Keynes, MK7 6AA, UK }

\begin{document}

\date{Accepted 2012 August 19}

\pagerange{\pageref{firstpage}--\pageref{lastpage}} \pubyear{2013}

\maketitle

\label{firstpage}

\begin{abstract}
\noindent This is the first paper in a series that present a multi-wavelength analysis of the archetype Ultra-Luminous InfraRed Galaxy (ULIRG) IRAS~FSC10214+47, a gravitationally lensed, starburst/AGN at $z=2.3$. Here we present a new lens model and spatially-resolved radio data, as well as a deep \hst\,F160W map. The lens modelling employs a Bayesian Markov Chain Monte Carlo algorithm with extended-source, forward ray-tracing. Using these high resolution {\sl HST}, {\sl MERLIN} and {\sl VLA} maps, the algorithm allows us to constrain the level of distortion to the continuum spectral energy distribution resulting from emission components with differing magnification factors, due to their size and proximity to the caustic. Our lens model finds the narrow line region (NLR), and by proxy the active nucleus, is preferentially magnified. This supports previous claims that preferential magnification could mask the expected polycyclic aromatic hydrocarbon spectral features in the {\sl Spitzer} mid-infrared spectrum which roughly trace the star-forming regions. Furthermore, we show the arc-to-counter-image flux ratio is not a good estimate of the magnification in this system, despite its common use in the IRAS~FSC10214+47 literature. Our lens modelling suggests magnifications of $\mu \sim 15-20\pm2$ for the \hst\,F814W, {\sl MERLIN}~1.7~GHz and {\sl VLA}~8~GHz maps, significantly lower than the canonical values of $\mu = 50-100$ often used for this system.  Systematic errors such as the dark matter density slope and co-location of stellar and dark matter centroids dominate the uncertainties in the lens model at the 40 percent level.
\end{abstract}


\section{Introduction}\label{section_introduction}

Over the past 15 years, both theory and observations have shown that star formation and the growth of super-massive black holes in galaxies are fundamentally linked \citep[e.g.][]{Silk1998,Magorrian1998,Ferrarese2000}. However, the physical mechanisms that drive the strong correlations of black hole mass ($M_{\rm BH}$) with both the stellar bulge mass ($M_{\rm bulge}$) and stellar velocity dispersion ($\sigma^{\star}$) are not yet clearly understood. Detailed observations of galaxies at cosmological distances play an essential part in the refinement of our theoretical and subsequent numerical models of galaxy evolution over cosmic time. High spatial resolution observations of galaxies with both starburst and AGN characteristics are able to give a unique perspective of the interplay of these components. This is a particularly powerful probe of galaxy evolution at z$\sim$2, corresponding to the peak of quasar and star formation activity \citep[e.g.][]{Dunlop1990,Madau1996}. However, detailed, high resolution studies of galaxies at this epoch are challenging as the imaging and 2D-spectral requirements are close to the limits of current instrumentation. To this end, many observations target gravitationally lensed sources which undergo a natural boost in flux and angular extent \cite[e.g.][]{Stark2008,Swinbank2010,Jones2010}. 

Lensing conserves surface brightness while increasing the apparent solid angle by a factor $\mu$. For unresolved sources, this results in a factor $\mu^2$ increase in effective sensitivity for a fixed observing time. The above characteristics make `cosmic telescopes' powerful tools in high-redshift astrophysics {\sl provided that} accurate lens models are derived. 

Although gravitational lensing is achromatic, different emission scales have differing magnification factors that can result in a distortion of the global spectral energy distribution (SED). This has been referred to as `differential' lensing by some authors, however we suggest this is an inappropriate term since any extended source is differentially magnified due to the optics of the system. We use the term `preferential lensing' for the purposes of this work to convey the idea that different emission regions (which dominate the global SED at different wavelengths) have different magnification factors based on their position and size. Therefore, preferential lensing at the very least avoids the confusion that arises between strong-lensing `apparent chromacity' and the magnification varying as a function of position for any narrowband image of a strongly-lensed system.

The effects of preferential lensing are often ignored or not quantified in studies that use lensing as a comic telescope to investigate the background source, especially in the (sub)millimetre regime where the spatial resolution is poor and/or multiple images are not detected and able to constrain the lens model. This paper presents first results of an extensive radio/mm observing programme to quantify this effect and the intrinsic starburst/AGN energy contributions in the famous strongly lensed system IRAS FSC10214+47 (IRAS~10214 hereafter). Our aim is not the study of the lens galaxy but rather to use it as a `cosmic telescope' in order to investigate the background source IRAS~10214.

\citet{Rowan-Robinson1991} first identified IRAS~10214 as the highest redshift (z $\sim$ 2.3) source detected in the IRAS Faint Source Catalogue. For a period of time in the early 1990s its apparent bolometric luminosity, ${L}_{\rm bol,app} \sim 10^{14} \ {\rm L}_{\odot}$, made it the most luminous object in the known Universe. This enormous luminosity, 99~percent of which is radiated in the infrared, led to suggestions that this was a primeval galaxy in the process of formation \citep{Rowan-Robinson1991}. However spectroscopic and morphological arguments purported IRAS~10214 to be gravitationally lensed by a foreground galaxy at redshift z $\sim$ 0.9 resulting in a total magnification between 50-100 at rest-frame optical/UV wavelengths \citep[see][]{Broadhurst1995,Serjeant1995,Close1995}, based on the arc-to-counter-image flux ratio. The lensing hypothesis was quite spectacularly confirmed by high resolution imaging with the {\sl Hubble Space Telescope} ({\sl HST}; \citealt{Eisenhardt1996}). Even after accounting for this high magnification factor, IRAS~10214 was, and still is, amongst the most luminous of the Ultra Luminous InfraRed Galaxies (ULIRGs).

Spectro-polarimetry revealed the rest-frame ultraviolet to be $\sim$28~percent polarised, as well as the presence of broad ultraviolet lines (\citealt{Goodrich1996}), providing strong evidence that this was the reflected light (presumably off dust clouds associated with the narrow-line region) from an obscured AGN. The source model of IRAS 10214 could now conclusively include an active nucleus, however X-ray observations with both the Chandra (\citealt{Alexander2005}) and XMM-Newton (\citealt{Iwasawa2009}) space telescopes made low S/N detections, seemingly inconsistent with the powerful AGN suggested by the luminous \oiii${\rm \lambda}$5007\AA\ line emission strength ($L_{\rm OIII} \sim$ 10$^{37}$ W, \citealt{Serjeant1998}). This lead to the suggestion that this is a Compton thick object (i.e. $N_{\rm H} > \sigma_{\rm T}^{-1} \, \sim 1.5 \times 10^{24} \, {\rm cm}^{-2}$). The lower than expected X-ray luminosity could also be explained by preferential lensing of the scattering clouds (i.e. the observable hot, ionised medium) due to a suggested offset from the AGN and as a result a larger magnification than that of the AGN. However, \citet{Nguyen1999} use arguments (that are re-investigated in \citealt{Deane2012c}) about the smooth polarisation position angle variation along the emission arc in their \hst ultraviolet map that limits this distance to $\sim$~40~--~100~pc in the source plane (assuming $\mu$ = 50 -- 100). The uncertainty in the intrinsic quasar bolometric luminosity at this stage was clearly very large given the suggestions of preferential lensing and/or Compton thick obscuration. 

Multiple detections of a large reservoir of molecular gas ($M_{\rm H_2}\sim10^{12} \, \mu^{-1}  \, $M$_{\rm \odot}$, \citealt{Brown1991}, \citealt{Solomon1992}, \citealt{Tsuboi1997}, \citealt{Radford1996}) supported suggestions that the far-infrared radiation originates from a massive starburst. Conditions of the inter-stellar medium were probed with the detection of higher transition CO lines (6-5, 7-6), as well as heavier atoms (e.g. \ci, \citealt{Ao2008}) and molecules found in the denser cores of star forming clouds (e.g. HCN (1-0), \citealt{VandenBout2004}). Despite this immense star-formation activity, mid-infrared {\sl Spitzer} spectroscopy did not reveal the strong polycyclic aromatic hydrocarbon (PAH) emission typical of ULIRGs \citep{Teplitz2006}. Furthermore, this same {\sl Spitzer} spectrum showed a 9.7 $\mu$m silicate feature in {\sl emission} as opposed to the usual silicate absorption seen towards obscured AGN. These attributes led \citet{Teplitz2006} to conclude that IRAS FSC10214 is unlike any ULIRGs or AGN in the local Universe. They proposed that the AGN is preferentially magnified and therefore masks the dominant starburst activity. From the very first publication on IRAS~10214 \citep{Rowan-Robinson1991} the question of the source of its enormous luminosity was highlighted: massive, dust-enshrouded star formation, or  a buried active galactic nucleus? Clearly this remains an outstanding question in the case of IRAS~10214 as well as a major challenge to current galaxy evolution models. 

High resolution radio imaging, not obscured by dust, offers a unique perspective on this galaxy which has composite (and in some cases counter-intuitive) properties across the electromagnetic spectrum, which are summarised in \S\ref{sec:overview}. In addition, resolving the various components within IRAS~10214 and their respective scales will allow an estimate of the level of preferential lensing in the system and its effect on the global SED. This apparent SED distortion has been invoked in the explanation of the peculiar features of IRAS~10214 \citep[e.g. ][]{Nguyen1999,Evans1999,Efstathiou2006,Teplitz2006,Ao2008} but it is difficult to explore without a lens model that has well defined uncertainties. Ironically, \citet{Matthews1994} argued that the observed chromacity in source structure suggested that IRAS~10214 was {\sl not} a lensed object. 

In this, the first paper in a series, our scientific goal is to derive a lens model for the IRAS~10214 system with well-defined uncertainties; and to use our spatially-resolved detections to derive magnification factors of each emission region probed. We wish to address the questions: can we decompose IRAS~10214 into different physical components with our resolved radio maps as well as our extended radio spectral coverage? If so, is a particular emission region preferentially magnified leading to the peculiar features found by the above authors? We include the \hst rest-frame ultraviolet map in this analysis for two reasons: (1) to derive a lens model with well-defined uncertainty; (2) this map is likely dominated by scattered quasar light meaning its source-plane and polarisation properties are valuable clues when addressing the above questions.

A more detailed study of preferential lensing is necessary as gravitational lenses provide us with the deepest views of galaxy formation and evolution in the Universe, and so any biases require investigation (e.g. effect on selection functions and the resultant source counts at various wavelengths, inferred star formation rate from high-redshift lensed CO detected galaxies, derived black hole masses of lensed sources, etc.). This is particularly relevant to the large sample of far-infrared selected, lensed galaxies discovered with the {\sl Herschel Space Observatory} \citep[e.g.][]{Negrello2010}, key to our understanding the fundamentally important population of `sub-millimetre' galaxies \citep[SMGs,][]{Blain2002} at $z \sim 2$, which are thought to be the progenitors of present day massive ellipticals. As new facilities come online in the next decade, the number of strongly-lensed systems will grow by orders of magnitude and so any systematic biases must be accounted for in detailed studies of individual systems. It will however become easier to do so with the superior sensitivity and angular resolution of forthcoming radio and millimetre facilities (e.g. {\it LOFAR, e-MERLIN, MeerKAT, JVLA, e-EVN, ALMA}) once fully commissioned and eventually with the {\it Square Kilometre Array (SKA)}. 

This paper is structured as follows: in \S2 we present the observations, including a deep \hst\,F160W map as well as resolved {\it MERLIN} and {\it VLA} radio maps. In \S3 we detail the lens model and Markov Chain Monte Carlo (MCMC) lens fitting technique. \S4 describes the results from the lens fitting, while in \S5 we synthesise all the results into a current interpretation of IRAS~10214 and close with conclusions and a future outlook in \S6. Throughout this paper we assume a concordance cosmology of $\Omega_{\rm M}$ = 0.27, $\Omega_{\Lambda}$ = 0.73, and $H_0$ = 71 km\,s$^{-1}$\,Mpc$^{-1}$ \citep{Spergel2007}, which yields an angular size scale of 8.3~kpc\,arcsec$^{-1}$ at the redshift of IRAS~10214 ($z$ = 2.2856, \citealt{Ao2008}).

\section{Observations}

The radio regime was one of the least studied sectors of the electromagnetic spectrum in the case of IRAS~10214 prior to this work. In this section we present the new and archival data that are used in this paper for both the lens model derivation and source plane investigation.

\begin{figure}
\includegraphics[width=0.43\textwidth]{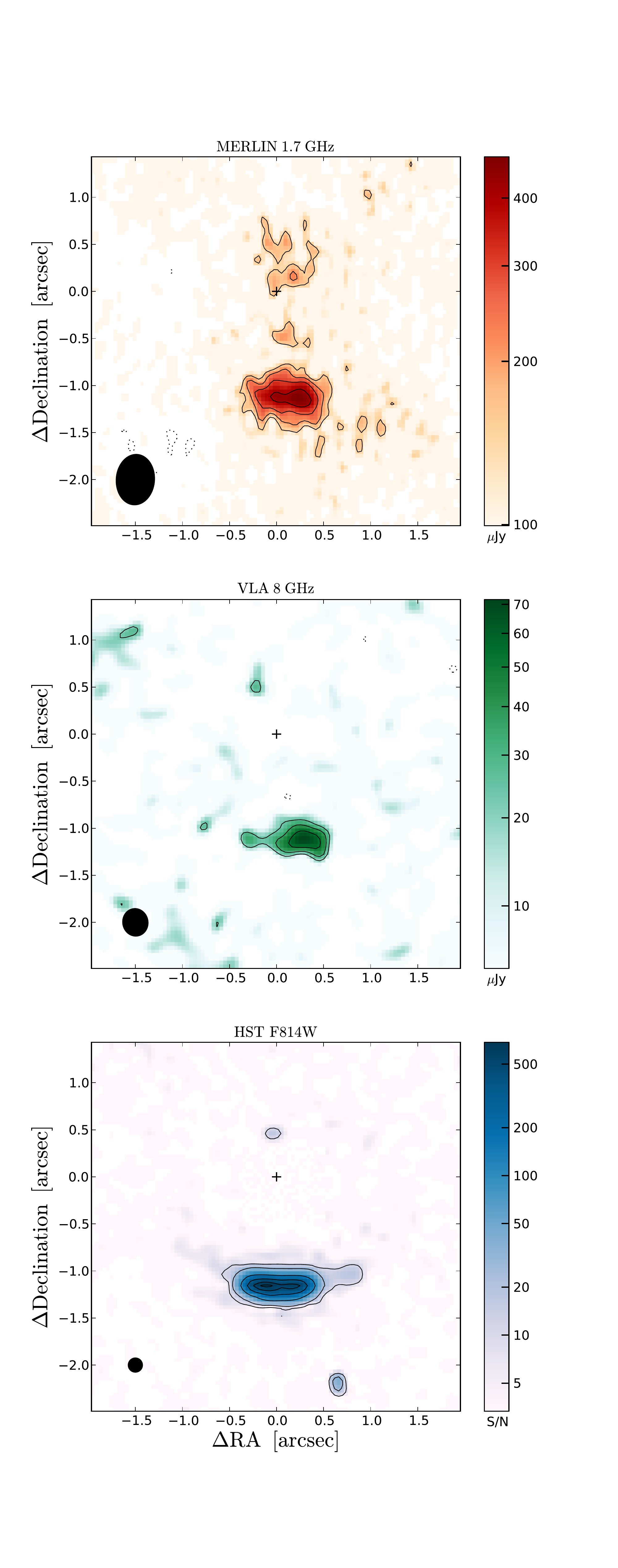}
\caption{{\bf Top panel:} \merlin 1.7~GHz map with $\sigma \, \sim$ 46 $\mu$Jy per 405 $\times$ 349 mas$^2$ beam. {\bf Middle panel:} 8~GHz \vla map with $\sigma \, \sim \,$11 $\mu$Jy per 292 $\times$ 267 mas$^2$ beam. Contours for both radio maps are at $\pm$3$\sigma$ and increase by a factor of $\sqrt{2}$. Dashed lines represent the negative contours. {\bf Bottom panel:} \hst\,F814W map with counter-image and arc. The integrated arc/counter-image flux density ratio  is $\check{\mu} \sim 75 \pm 25$. The \hst PSF through this filter has a FWHM of $\sim$100 mas, as detailed in \citep{Eisenhardt1996}. Note that the lens has been fit using GALFIT and removed from this image. The FWHM of all PSFs  are shown in the lower left of each frame. In all panels the cross indicates the Sersic-fitted centroid of the lensing galaxy as measured from the \hst\,F160W map which is within 10~mas of the equivalent measurement with the \hst\,F814W map. }
\label{fig:maps}
\end{figure}

\subsection{Archival Observations}

A significant number of competitive, yet unpublished observations lay in the {\sl MERLIN}, \vla and \hst archives prior to this work. We obtained all these datasets and reprocessed them as is described below. We also use the quoted flux densities and centroid positions of the \vla 1.49~GHz, 4.86~GHz and 8.44~GHz maps described in \citet{Lawrence1993}. 

\subsubsection*{{\bf \it MERLIN} 1.7~GHz}\label{sec:1.6ghz}

IRAS~10214 was observed for 24 hours with the \merlin array on 2 November 1995. The observation included the 76 metre Lovell Telescope as well as the Wardle antenna making this an 8 element observation (28 baselines). {\sl J}\,1027+480 was used as a phase calibrator (angular separation $\Delta \theta \sim 1^\circ$) and {\sl JVAS J}\,2136+0041 was used as a pointing calibrator (positional accuracy $<$ 0.36 mas). IRAS~10214 was observed for an average of  5~minutes in the roughly 7.5 minute phase-target cycle (this period varied by $\sim$30~percent over the course of the observation). Preliminary calibration of the {\sl uv}-data set was carried out with the {\sl MERLIN} {\sc aips} pipeline\footnote{http://www.merlin.ac.uk/user\_guide/AIPS\_scripts.html}, which performs initial fringe-fitting (calibration of delays, rates and phase), and phase and amplitude calibration. Following this preliminary calibration, a more detailed, manual calibration is performed in {\sc apps}. This was performed in a cyclic process with detailed data editing to excise Radio Frequency Interference (RFI); refine the delay calibration, as well as perform phase and amplitude self-calibration on the phase reference source.

We achieve a 1-$\sigma$ noise sensitivity of $\sim$46~$\mu$Jy per 405 $\times$ 349 mas$^2$ naturally-weighted beam (Briggs {\tt robust} weighting parameter = 5). The resultant map reveals a resolved `arc' (Fig.~\ref{fig:maps}, top panel) roughly co-spatial with the \hst\,F814W image \citep{Eisenhardt1996}, however the \emph{dominant} peaks of these two maps are offset by $\sim$0.4~arcsec in RA. A resolved 1.7~GHz detection is only achieved for visibility weightings with a Briggs {\tt robust} weighting parameter $>$ 2. 
The foreground lensing galaxy centroid is indicated by a cross in all three panels in Fig.~\ref{fig:maps}. The 1.7~GHz map shows a number of 2-3~$\sigma$ peaks that are seen in this region as well as the counter-image position as reported by \citep{Eisenhardt1996}, however nothing further can be reliably inferred at this low significance. To probe larger scale features we taper the {\it uv}-data ($ < \, 700 \ \mathrm{k} \lambda$) which imposes a Gaussian weighting to the data as function of baseline length measured in wavelengths. This is a function centered on 0 $\lambda$ and has a 700 k$\lambda$ FWHM. 

A comparison of the {\sl uv}-tapered and full resolution maps suggests that no flux is resolved out within the uncertainties. In addition, the 330 MHz through 4.8~GHz apparent spectral index suggests, within the errors, that no significant extended flux is resolved out in the \merlin 1.7~GHz observations. It appears most likely that this 2-3~$\sigma$ emission is associated with the lens given its inconsistency with the plausible counter-image locations. Radio emission at the mJy level in lensing galaxies is not unexpected, as shown in a CLASS\footnote{Cosmic Lens All-Sky Survey, \mbox{www.aoc.nrao.edu/$\sim$smyers/class.html}} example \citep{More2008} as well as a sample of 10 SDSS\footnote{Sloan Digital Sky Survey, www.sdss.org}-selected lenses \citep{Boyce2006}.

\subsubsection*{\vla A-array 8~GHz}

IRAS~10214 was observed twice with the \vla (A-array) in the 8~GHz band ({\sl X}-band) in the 1990s, both with consistent flux densities and centroid positions. The second, deeper observation (unpublished) achieved a sensitivity of 13~$\mu$Jy\,beam$^{-1}$ after careful calibration and editing. Both datasets were re-calibrated using standard techniques in the {\sc aips} data reduction package, following which the two calibrated {\sl uv}-datasets were combined using the {\sc DBCON} task. These observations were not at the typical frequencies where antenna gain curves are required ($>$ 15 GHz), nonetheless they were applied to maximise accuracy. The re-reduced 1991 {\sl uv}-dataset was precessed from Equinox B1950, Epoch B1979.0 (barycentric) to J2000 coordinates with offsets calculated with the STARLINK {\tt COCO} task set on high resolution (0.1 mas precision). The peak of the precessed 1991 map is co-spatial with the 1995 map. Astrometric accuracy at the \vla can be $\Delta \theta \simeq 50$\,mas  in A-array in good conditions and $\Delta \theta \simeq 100$ mas under more normal conditions (Ulvestad~et~al. 2009). The absolute positions of the two phase calibrators used in the 1995 observation ({\sl J}\,1033+4116 and {\sl J}\,1027+480) were within $\Delta \theta <$ 3 mas and 31 mas of their catalogued positions respectively. The positional errors are ($\Delta$RA = 0.20 mas, $\Delta$Dec = 0.40 mas) and ($\Delta$RA = 0.45 mas, $\Delta$Dec = 0.75 mas), as sourced from the International Celestial Reference Frame (ICRF, \citealt{Ma1998}) and the \vlba Calibrator Survey (VCS1, \citealt{Beasley2002}) respectively. The absolute position of the phase calibrator used in the 1991 observation ({\sl B}\,1031+567) was within $\Delta \theta <$ 5 mas of its catalogued position. The positional error is $\Delta$RA = 0.31 mas, $\Delta$Dec = 0.47 mas, sourced from the ICRF. The radio coordinate reference frame must be directly compared to the optical reference frame, which has been done by \citet{Lawrence1993} who compare the optical and radio positions of 20 compact radio sources in the region around IRAS~10214. They find no significant mean difference, however, they note the error on the mean difference is 0.2~arcsec which we take as the systematic uncertainty in the radio-optical co-ordinate system alignment. Since we use the same data, and our calibrators (as well as the 8~GHz centroids) are consistent, we assume have an equivalent astrometric matching between radio and optical reference frames. 
 
 The {\sl uv}-datasets were combined and transformed to produce the map in Fig.~\ref{fig:maps} (middle panel). Apart from the detection of the main peak associated with the optical arc, our combined 1991/1995 8~GHz map (Fig.~\ref{fig:maps}, middle panel) reveals two lower significance ($\sim$3-$\sigma$) peaks, one eastward of the optical arc, the other eastward of the optical counter-image. The latter has a very low arc/`counter-image' ratio ($\check{\mu}$~=~8), inconsistent with all previous lensing results (and this work, as seen in \S\ref{sec:results}). The `counter-image' is not seen in either of the 1991 and 1995 maps and so despite its 4-$\sigma$ significance this is not a robust feature. We therefore assume the detection is {\sl not} secure leading to an 8~GHz lower limit of $\check{\mu} \, > \,$24, however the 4$-\sigma$ peak remains in the lens fitting procedure allowing the Markov Chain Monte Carlo algorithm described in Section~3 to test whether this is a feasible counter-image.

\subsubsection*{\hst\,F814W}

\begin{figure}
\centering
\includegraphics[width=0.45\textwidth]{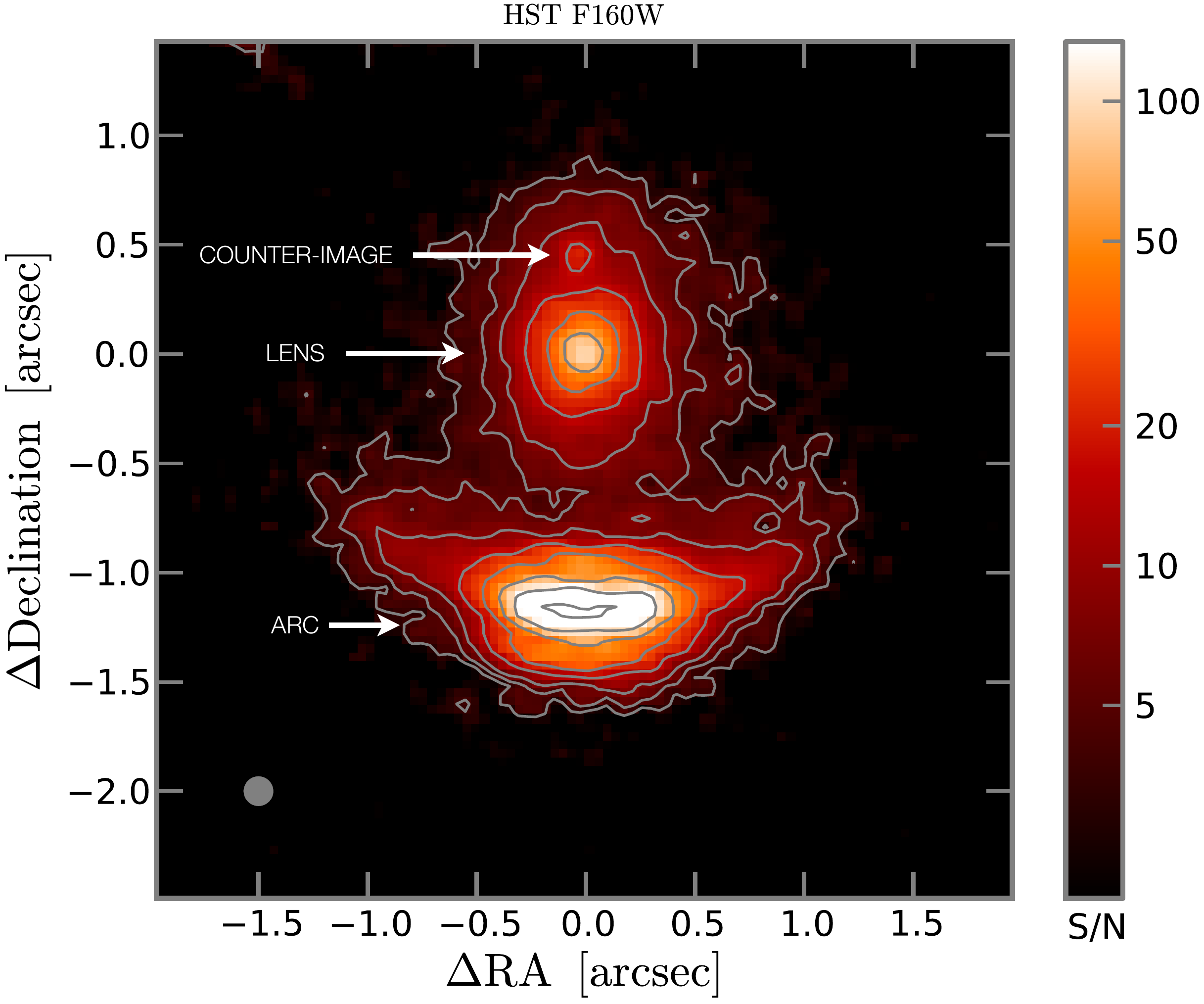}
\caption{ \hst\,F160W map with lensing galaxy, counter-image and the arc. The integrated arc/counter-image flux density ratio is $\check{\mu} \sim$69 $\pm$7. The HST PSF through this filter has a FWHM of $\sim$150 mas which is illustrated by the grey circle in the lower left. Note that the colour-scale is logarithmic to enhance the extended low surface brightness arc.}
\label{fig:hst160}
\end{figure}

IRAS~10214 was observed with the WFPC2 F814W filter during two orbits on 10 December 1994. This rest-frame ultraviolet map was the first conclusive evidence that IRAS~10214 is indeed gravitationally lensed. All data were processed through the Canadian Astronomy Data Centre (CADC) \hst pipeline\footnote{http://cadcwww.dao.nrc.ca/hst}. The morphology (see Fig.~\ref{fig:maps}, bottom panel) shows a clear $\sim$1~arcsec long arc and a low S/N detection of the counter-image. The lensing galaxy has been subtracted using a GALFIT 3 \citep{Peng2010} Sersic fit. The integrated arc/counter-image flux density ratio  is $\check{\mu} \sim 75 \pm 25$. The observations, derived properties and lens modelling are fully described in \citet{Eisenhardt1996}. Two key attributes of this map are accurate point spread function (PSF) characterisation and astrometry. The PSF FWHM is $\sim$100 mas as derived from the `Tiny Tim' \hst software package using a $K$-star source colour since an accurate empirical estimate was not possible. This is fully described in \citealt{Eisenhardt1996} and is consistent with two stars in the field, one of which is saturated and the other too weak for accurate PSF estimation. The astrometry is in agreement with determinations from \citet{Nguyen1999,Eisenhardt1996,Evans1999} and Simpson~et~al.~(in preparation) to within $\Delta \theta <$ 10 mas.

\subsubsection*{\hst\,F160W}

IRAS~10214 was observed with the NICMOS camera 2 on 22 March 2004 within a single orbit under Proposal ID 09744 (PI: Kochanek). The wide band filter F160W was used, resulting in a PSF FWHM of $\sim$0.15~arcsec. The map (Fig.~\ref{fig:hst160}) shows a strong arc structure, the lens as well as a counter-image to the north of the lens. The arc shows two clear components: an extensive, faint arc, as well as a larger, more dominant component. We attribute the latter to the scattered quasar light (that dominates the \hst\,F814W map) as well as a more extended, lower magnification host galaxy component. This is supported by the global SED as well as the 4000 $\AA$ break first identified by \citet{Lacy1998}. The detection of the counter-image has higher significance than the \hst\,F814W map and an arc/counter-image flux ratio $\check{\mu}$ = 69 $\pm$ 7. Despite this higher S/N we do not use the \hst\,F160W map for our lens modelling since it clearly has a complex source structure. A larger version of the \hst\,F160W map (Fig.~\ref{fig:hst160}) is shown later (Fig.~\ref{fig:satgals}). For the purposes of this work, the larger field-of-view map is used to investigate the lensing potential of nearby galaxies (\S\ref{sec:systematics}) and to determine the ellipticity and position angle priors for main lens (\S\ref{sec:lensmodel}). The source plane properties of the \hst\,F160W map will be shown in Deane~et~al.~(in preparation). All data were processed through the CADC \hst pipeline and astrometry was corrected to be consistent with \citet{Nguyen1999}, \citet{Eisenhardt1996}, \citet{Evans1999} and Simpson~et~al.~(in preparation), as in the case of the \hst\,F814W map.

\subsection{New Observations}

In order to increase the radio spectral coverage of IRAS~10214, we performed unresolved low and high frequency observations with the \gmrt and \ami arrays respectively. 

\subsubsection*{GMRT 330 MHz}

\begin{figure}
\includegraphics[width=0.45\textwidth]{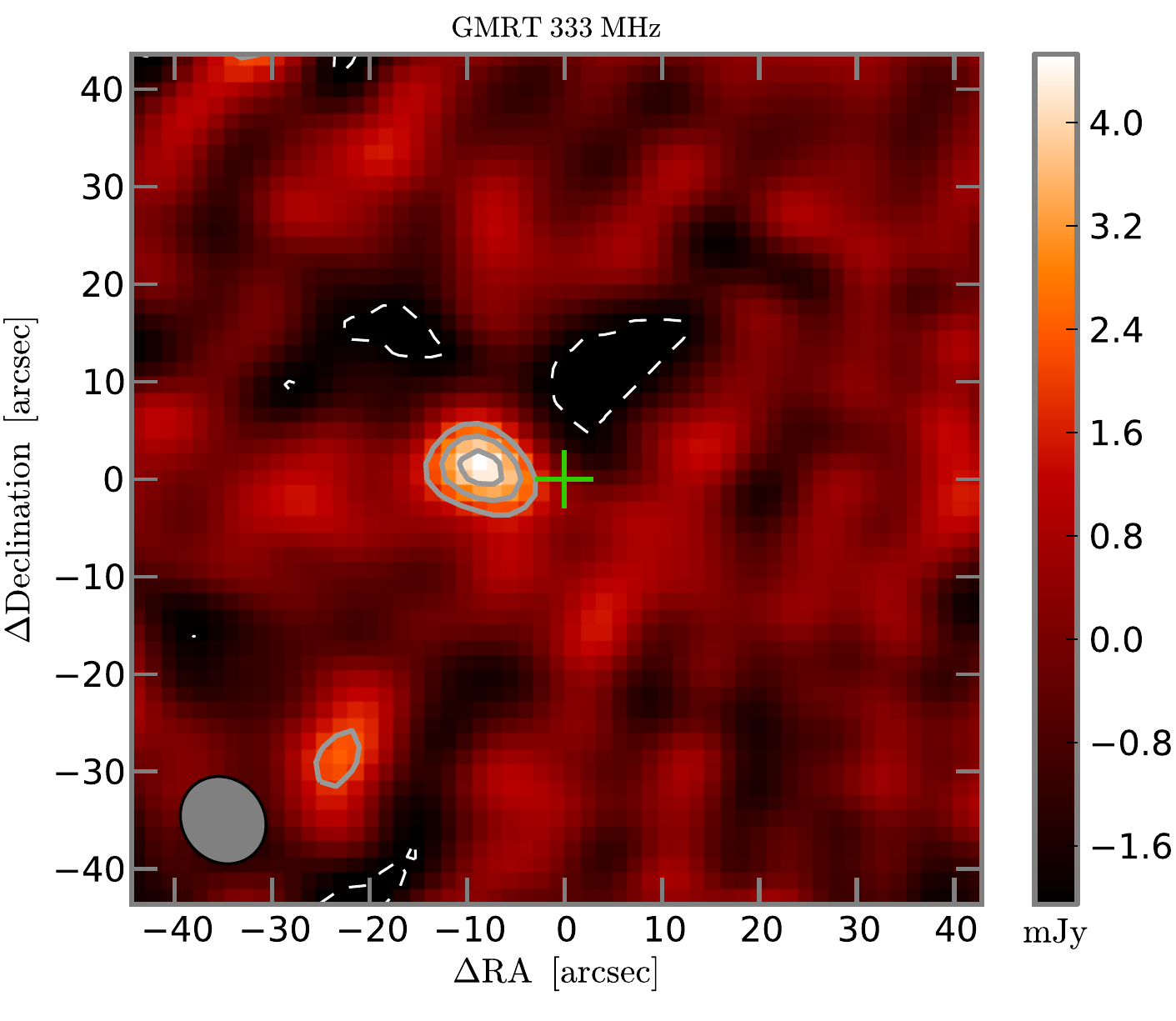}
\caption{{\sl GMRT} map with grey contours starting at $\pm$3-$\sigma$ in 1-$\sigma$ steps of $\sigma \sim \,$0.71 mJy per $9.7\, \times \,$7.8 arcsec$^2$ beam. The integrated flux density is $S_{\rm int}  \sim 4.42 \pm 1.48$~mJy. The green cross indicates the position of the centre of the arc of IRAS~10214 (RA = 10$^{\rm h}$ 24$^{\rm m}$ 34.564$^{\rm s}$, Dec = 47$^{\circ}$ 09' 9.61''). Note that the size of this cross is purely illustrative and does not represent the astrometric uncertainty of $\Delta \theta \sim 1$~arcsec for the \gmrt detection. Absolute astrometric accuracy is compromised in the faceted self-calibration process which is essential to achieve reasonable dynamic range over the large ($>$ 1 deg$^2$) field of view at low frequencies.   }
\label{fig:GMRTmap}
\end{figure}

\begin{figure}
\includegraphics[width=0.45\textwidth]{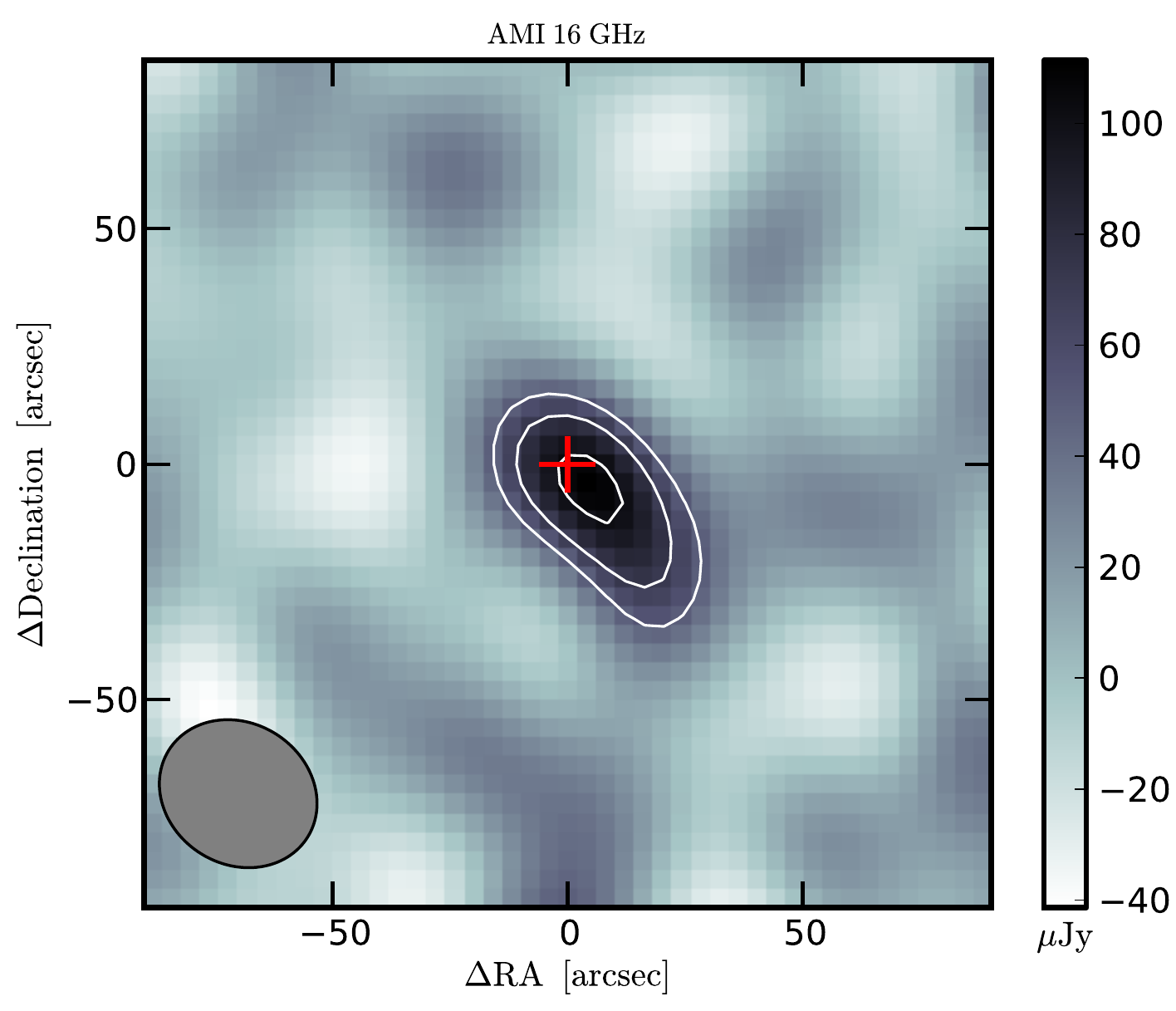}
\caption{AMI map with white contours starting at $\pm$3-$\sigma$ in 1-$\sigma$ steps of $\sigma \sim \,$17~$\mu$Jy per 34.8$\, \times \,$30 arcsec$^2$ beam. The integrated flux density is $S_{\rm int} \, \sim \,$144~$\mu$Jy. The red cross indicates the position of the centre of the arc of IRAS~10214 (RA = 10$^{\rm h}$ 24$^{\rm m}$ 34.564$^{\rm s}$, Dec = 47$^{\circ}$ 09' 9.61''). Note that the size of this cross is purely illustrative and does not represent the astrometric uncertainty of $\Delta \theta \sim 2$ arcsec.   }
\label{fig:maps16ghzmap}
\end{figure}

\noindent IRAS~10214 was observed with the \gmrt array on 3 July 2009 at
330~MHz for 25 minutes. Data processing was performed with an automated calibration
and imaging pipeline. The pipeline is based on {\sc python, aips} and
{\sc parseltongue} \citep{Kettenis2006} and has been specially
developed to process \gmrt data. The full technical description of this calibration and imaging pipeline is extensive, and therefore presented in detail in Kl\"ockner~(in preparation). Here we just give a brief description of the amplitude calibration.  Absolute amplitude calibration is based on the {\sc AIPS} task {\sc
SETJY}. Instead of using a pseudo-continuum measure to perform 
amplitude calibration, a single channel (chan \#60, $\nu$ = 332.5 MHz) is used for improved accuracy. This 
resulted in a flux density measurement of 46.07~Jy for 3C~147. The final $\sim$1.4~deg field-of-view image is catalogued and evaluated on the basis
of known radio catalogues such as FIRST, NVSS, WENSS \citep{Becker1995,Condon1998,Rengelink1997} in order to verify
the flux density calibration and astrometric accuracy (Mauch et al., in preparation). 

\noindent The maximum in the final image is 706~mJy~beam$^{-1}$ and the noise
is 0.56~mJy\,beam$^{-1}$ at an averaged observing frequency of 333~MHz. The
global dynamic range, using the maximum flux density and the noise, is
$\sim$1253, whereas the local dynamic range, which is determined at 7
bright sources in the field and is 107. At a 5-$\sigma$
detection limit of 2.82~mJy, 262 sources were catalogued.

IRAS~10214 is automatically detected
at 6-$\sigma$, based on the local noise estimate quoted in Table~\ref{tab:fluxes}. The source is not extended (see Fig.~\ref{fig:GMRTmap}), having a fitted major- and
minor-axis of 9.7$\times$7.8~arcsec and a position angle of 60$^{\circ}$, the
uncertainty in the positional fitting is 0.2~arcsec.  The flux
density measures of IRAS~10214 at 333~MHz are the peak flux density
$S_{\rm peak}$~=~4.42$\pm$0.71~mJy\,beam$^{-1}$ and integrated flux density $S_{\rm int}$ = 
4.42$\pm$1.48~mJy.

\subsubsection*{16 GHz Arcminute Micro-Kelvin Imager Observations}

IRAS~10214 was observed on 15 December 2009 for 12 hours by the 
AMI-LA~\citep{Zwart2008}.
The telescope observes between 13.5 and 18~GHz in six spectral
channels of 0.75~GHz bandwidth. The source {\sl J}\,1027+4803 was observed for 100~s every 600~s
for accurate phase calibration. Flux calibration was performed using short observations
of 3C\,286, which was assumed to have an I+Q flux of 3.53~Jy at 15~GHz
and a spectral index of 0.72~(Perley, private communication). The data
were flagged and calibrated using {\sc reduce}, a local software
package developed for AMI. A total of 14\% of the data were flagged
due to a combination of pointing errors, shadowing and RFI.  The
calibrated data were then mapped and {\sc clean}ed in {\sc aips} using
the {\sc imagr} task.  The integrated flux density of the detection in Fig.~\ref{fig:maps16ghzmap} was
$S_{\rm int} \, \sim \,$144 $\mu$Jy. 
The final thermal noise on the continuum
map was $17 \, \mu$Jy\,beam$^{-1}$ and the map has not been corrected for the
telescope primary beam response, which is well modelled by a Gaussian
of FWHM 5.6~arc-minutes. The {\sc clean} restoring beam is $34.8
\, \times \,  30.0$~arcseconds at a position angle of 32.2~degrees. We do not expect confusion with 
other sources within this beam based on the higher resolution 8~GHz \vla continuum map as well as 
 a 35 GHz \evla continuum map with 1-$\sigma \sim 40 \ \mu$Jy\,beam$^{-1}$ \citep{Deane2012b}.

\begin{table}
\centering
\scriptsize
\begin{tabular}{c  c  c  c  c  c }
\hline
{\bf Telescope} &  {\bf Frequency}  &  {\bf Flux Density$^{\dagger}$}  &  {\bf Noise}  &  {\bf Bandwidth}  &  {\bf Beam} \\ 
                &   GHz             &       mJy            &     $\mu$Jy\,beam$^{-1}$ &   MHz       &   arcsec$^2$  \\
 \hline \hline 
\gmrt            &   0.33            &     4.42  $\pm$1.48            &    706          &      16       &  9.7 $\times$ 7.8          \\
\merlin            &   1.66            &     1.12   $\pm$0.12        &    46          &      13       &  0.41 $\times$ 0.35          \\
\vla             &   8.4            &     0.28 $\pm$0.03             &    11          &      100       & 0.29 $\times$ 0.27          \\
\ami             &   15.75            &     0.144 $\pm$0.032             &    17          &      4500       & 34.8 $\times$ 30.0         \\ 
\hline
\end{tabular}
   \caption{Summary of all new radio observations of IRAS~10214. \newline
   $^\dagger$ Integrated flux densities for the unresolved maps ({\sl AMI, GMRT}) are from a fitted Gaussian, the uncertainty of which is listed alongside. Flux densities from the resolved maps ({\sl MERLIN}, \vla) are calculated by summing over a region defined by a 2.5-$\sigma$ threshold clip, while the uncertainties are defined by the quadrature sum of the absolute amplitude calibration uncertainty and the naturally-weighted map sensitivity.}
   \label{tab:fluxes}
\end{table}

\subsection{Multi-wavelength Overview }\label{sec:overview}

\begin{figure*}
\begin{center}
\includegraphics[width=1\textwidth]{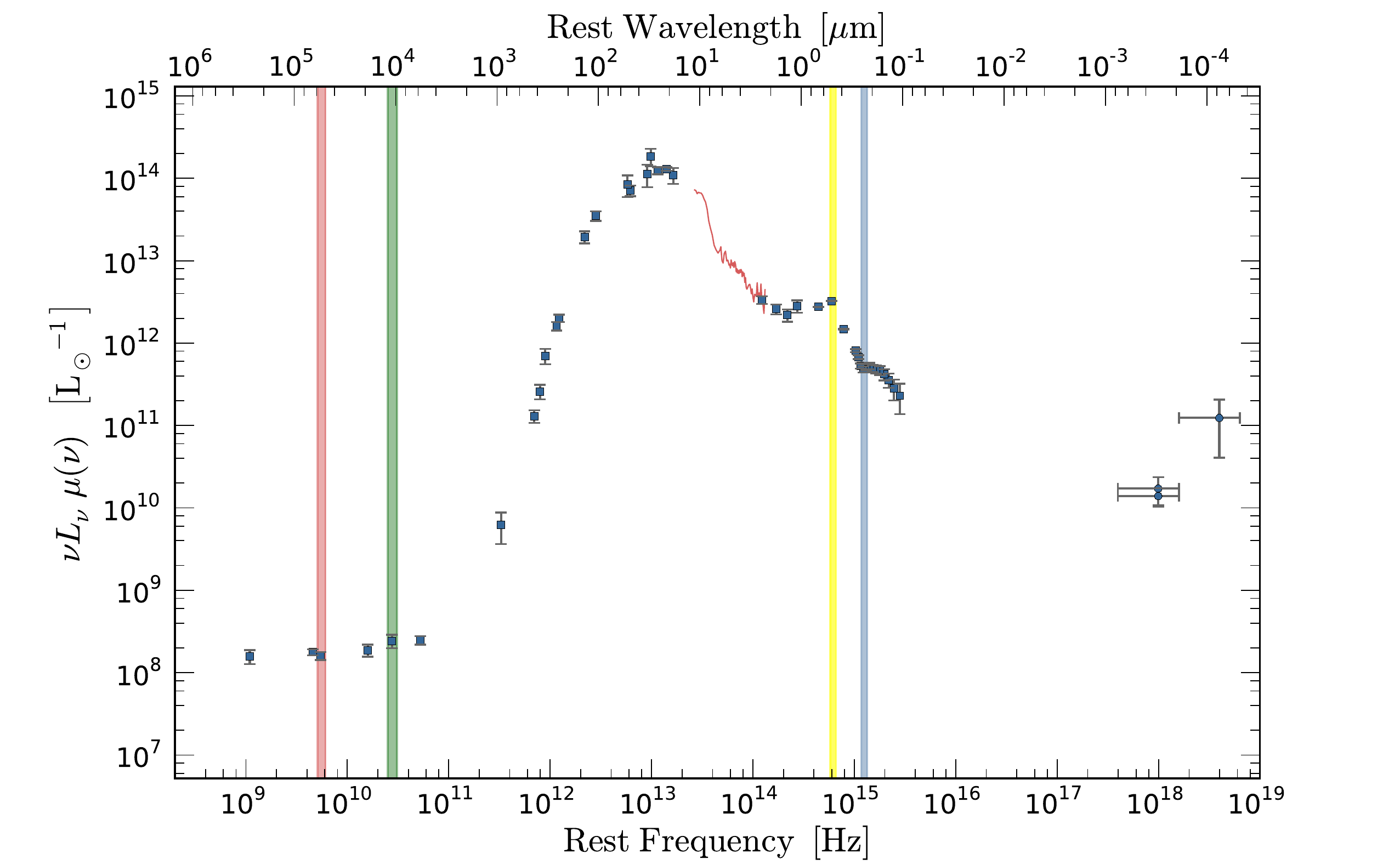}
\end{center}
\caption{The \emph{apparent} (i.e. lensed) X-ray through radio SED of IRAS FSC10214 showing the three bands probed by the high resolution imaging presented in this work: radio lobe (red), \vla~8~GHz (green), scattered quasar light (blue) as traced by the \merlin 1.7~GHz, \vla 8~GHz and \hst WFPC2 instruments respectively. The yellow band represents the \hst NICMOS 1.6 $\mu$m (F160W filter) apparent luminosity which appears to trace the host galaxy's stellar component. The dark red line in the mid-infrared is the {\sl Spitzer} spectrum \citep{Teplitz2006}. Note that the parameter $\mu(\nu)$ accounts for the magnification boost which varies as a function of frequency due to differing sizes and positions of emission components. This function is as yet undetermined and hence included in the ordinate label. The flux densities used in this plot are sourced from \citet{Rowan-Robinson1991,Downes1992,Rowan-Robinson1993,Lawrence1993,Barvainis1995,Benford1999,Teplitz2006,Alexander2005,Solomon2005,Ao2008,Iwasawa2009}.}
\label{fig:sed}
\end{figure*}

To first order, IRAS~10214's global SED (Fig.~\ref{fig:sed}) is typical of an obscured AGN with significant star formation \citep[e.g.][]{Martinez-Sansigre2005}. Both of these properties are confirmed by a number of spectral features including a polarised ($>$20\,\%), broad ($\sim$6000 km\,s$^{-1}$) \civ\ line \citep{Goodrich1996} and a vast reservoir of molecular hydrogen ($\mathrm{H}_2 \sim 10^{11}~\mu^{-1}~\mathrm{M}_{\odot}$,  \citealt{Ao2008}, \citealt{Solomon1992N}). There is substantial mid- to far-infrared dust emission ($M_{\rm dust} \, \sim \, 5\times 10^9 \, \mu^{-1} \ \mathrm{M}_{\odot}$), strong NIR emission with a steep downward slope toward shorter wavelengths which suggests high extinction by a dusty nuclear toroidal structure (under the unified quasar model, \citealt{Antonucci1993}) and/or the host galaxy \citep[see ][]{Martinez-Sansigre2005}. 
The host galaxy appears to dominate in the optical band (rest-frame) as shown by the 4000 \AA\ break observed in the NIR spectrum of \citet{Lacy1998}. 

Our extended radio SED coverage appears to confirm the steep spectrum component, first measured in \citealt{Rowan-Robinson1993} ($\alpha = 0.92 \pm 0.04$, where \begin{math} \alpha \equiv   - \log(S_1 / S_2) / \log( \nu_1 / \nu_2) \end{math}) between $\nu_{\rm obs} \sim$330 MHz to 4.8~GHz. The spectral index uncertainty is Monte Carlo derived, incorporating the flux density uncertainties from four measurements in this frequency range. The spectrum flattens between 4.8~GHz and 16 GHz with an apparent peak at 8~GHz, plausibly due to the presence of a flat-spectrum radio core. However, the flux measurements at 8~GHz and 16~GHz are separated by $\sim$15 yr, which increases the spectral index uncertainty due to instrinsic quasar variability. Nonetheless, our measured spectral indices are $\alpha^{8.4}_{4.8} = 0.53  \pm 0.09\, (0.24)$ and $\alpha^{16}_{4.8} = 0.78 \pm 0.17 \, (0.26)$, where the uncertainties are Monte Carlo derived and those quoted in parentheses include the absolute flux calibration uncertainties. 

 Inspection of the maps in Fig.~\ref{fig:maps} reveals a spatial offset of $\sim$0.4$\pm 0.1$ arcsec between the 8~GHz peak and both the dominant \hst\,F814W peak and the quoted \vla 1.4~GHz centroid in \citet{Lawrence1993}. This offset of the 8~GHz peak is present in both the 1991 and 1995 8~GHz datasets. Adding the astrometric errors from the 1991 1.49~GHz and 8.44~GHz observations in quadrature, \citet{Lawrence1993} find an offset between the optical/NIR positions and the average radio position $\Delta \theta = 0.63 \pm 0.37$, consistent with our more sensitive observations here. Therefore, this offset appears reasonably robust and independently verified. 

The regions of the global SED probed by the spatially-resolved maps presented in this work are illustrated by the vertical lines on the SED plot in Fig.~\ref{fig:sed}: \merlin 1.7~GHz (red), \vla 8~GHz (green), \hst\,F814W (blue), and \hst\,F160W (yellow). The \hst\,F160W filter appears to be dominated by stellar emission based on the SED shape, and the 4000~\AA\ break \citep{Lacy1998}, whereas the \hst\,F814W map of \citet{Eisenhardt1996} is dominated by scattered quasar light as inferred from \hst polarisation observations. 

In subsequent modelling, we assume that the 1.7~GHz map is dominated by a radio lobe. The term `radio lobe' is used relatively loosely here: the steep apparent spectral index suggests that this is an aged distribution of relativistic electrons in a magnetised plasma. As we will see in \S\ref{sec:results}, the inferred source plane brightness temperature is evidence that this is not the core of a high powered jet. We therefore describe it as a radio lobe with the full expectation that some of this non-thermal emission is associated with star formation. While the radio spectrum flattening around 8~GHz suggests that this map may be dominated by a radio core, we show in \citet{Deane2012c} that VLBI observations are not consistent with this. We therefore expect the 8~GHz map includes emission contributions from star formation and/or radio jets. We emphasize here that we are modelling emission regions and not physical components, however, the polarisation properties of the \hst\,F814W map suggests that this region in particular is dominated by a single physical component (i.e. the scattered quasar light). Finally, we later argue that the majority of the quasar light scattering occurs in the NLR. Therefore, the \hst\,F814W map will also be used as a rough tracer of the NLR.

\section{Lens Modelling}\label{sec:lensmodel}

Gravitational lensing occurs as a result of the apparent deflection of light between the astrophysical source and observer by an intervening mass that curves the local spacetime. The gravitational lensing formalism is comprehensively covered in many texts \citep[e.g.][]{Blandford1992,Schneider2006,Treu2010} and therefore, in the interests of brevity, is not repeated here.

\subsection{Lens Model Derivation}

We derive the lens model parameters, given the data which are the \hst\,F814W map pixel values presented in Fig.~\ref{fig:maps} (bottom panel). This is not only a very high S/N map but is at a wavelength we believe to be dominated by a single physical source (i.e. the emission is dominated by scattered QSO light, despite showing some structure along the arc) and therefore likely to have a simpler intrinsic source structure than a filter with both stellar and QSO components. We predict data pixel values which are derived by ray-tracing trial source models into the image plane based on a trial lens model. We have developed software that implements a MCMC routine to compare ray-traced trial models to the data and so sample the posterior probability distribution function (PDF) of all free parameters. The posterior PDF is defined by Bayes' Theorem

\begin{equation}
Pr(\bs{x} | \bs{d},H)  =   \frac{Pr(\bs{d} | \bs{x},H) \,  Pr(\bs{x} | H)}{ Pr(\bs{d} | H)},
\end{equation}

\noindent where $Pr(\bs{d} | \bs{x},H)$ is the likelihood and $Pr(\bs{x} | H)$ is the product of our informative priors to be described below. $Pr(\bs{d}|H)$ is known as the \emph{evidence} which normalises the posterior. In our MCMC routine we sample the unnormalised posterior distribution.  All ray-tracing from source to image plane is performed by routines in the software package {\sc glamroc}\footnote{Gravitational Lens Adaptive Mesh Raytracing of Catastrophes, see http://kipac.stanford.edu/collab/research/lensing/glamroc}. These ray-tracing routines adaptively increase the spatial resolution (i.e. the number of pixels) as a function of the local magnification, the thresholds of which can be set by the user. This improves the accuracy (and/or computational efficiency) when modeling lensed sources that have close proximity to the caustic. This software is especially powerful in the cusp-caustic lens configuration as in the case of IRAS~10214. Computational efficiency limits the source plane resolution, however the magnification around the cusp can change dramatically inside a single source plane pixel. Therefore, {\sc glamroc} alleviates the computational challenge by enabling higher resolution in regions that exceed a defined magnification. This sub-division of source plane pixels can be repeated an arbitrary number of times.

 Before describing the MCMC routine in further detail, we first introduce the existing IRAS~10214 lens model and describe the observables on which we base our informative priors. The most sophisticated IRAS~10214 lens model in the literature is that derived by \citet{Eisenhardt1996}. They use a least-squared approach, minimising the distance between the model and the brightest 96 pixels in the arc as well as the brightest pixel in the counter-image, using the \hst\,F814W data. They assume a uniform (constant surface brightness) circular disk source model and found a best-fit isothermal ellipsoid mass distribution ellipticity of $\varepsilon$~=~0.3. The observed stellar ellipticity of the lensing galaxy is $\varepsilon \, \sim \,$  0.16-0.2 for the \hst 0.8, 1.1 and 1.6 $\mu$m images. In the Sloan Lens ACS Survey (SLACS), \citet{Koopmans2006} model 15 massive early-type field galaxies with a singular isothermal ellipsoid (SIE). They find that the stars and dark matter have highly correlated position angles ($PA$) and ellipticities ($\varepsilon$). The average ratio of dark matter to stellar ellipticity is $\langle q_{\rm SIE}/q_* \rangle = 0.99 \pm 0.11$, and the average difference between the derived dark matter $PA$ and stellar $PA$ is $\langle \Delta PA \rangle = 0^{\circ} \pm 10^{\circ}$, the uncertainties indicate the total scatter $\sigma$, not $\sigma / \sqrt{N}$. Leveraging off this evidence that baryons in lensing galaxies broadly follow the dark matter potential, we select an observationally motivated approach and set lens model parameter priors based on the baryon distribution. The \hst\,F160W observation offers two major advantages in the gravitational lensing analysis of IRAS~10214. 

\begin{enumerate}
\item {\bf Sensitivity:} the \hst\,F160W map has roughly a factor of 2 better S/N than the \hst\,F814W map. 
\item {\bf Dark Matter Tracer:} the foreground lens galaxy is observed at rest-frame {\sl R}-band tracing the older, more virialised stellar population which should, in principle, better sample the dark matter potential. 
\end{enumerate}

We therefore use this \hst\,F160W map to set the lens model priors. We derive the {\em mean} of these position angle and ellipticity priors using GALFIT two-dimensional Sersic fits. GALFIT is run for $5 \, \times \, 10^4$ Monte Carlo iterations with each initial parameter randomly varied by an order of magnitude (or between 0.5-8 for the Sersic index, 0--180$^{\circ}$ for the position angle). 
We find an ellipticity  $\varepsilon \, = 0.20 \, \pm_{0.004}^{0.01}$ at a  position angle $\theta_{\rm PA}$ = -4.8$^{\circ} \, \pm \, 0.9^{\circ}$ (East of North) and a Sersic index\footnote{Note that the lens is simultaneously fit with a Sersic component and an unresolved core component. The luminosity of this core component contributes $\sim$ 0.4\% of the total and therefore does not make a significant contribution to the mass within the Einstein radius. } $n$ =  2.65 $\pm$0.23, in the range between that measured for typical ellipticals ($n = 4$) and disk galaxies ($n = 1$).

These results are used to set Gaussian mass distribution priors of $\varepsilon_{\rm prior}$ = 0.2 $\pm$0.4 and $PA_{\rm prior}$ = -4.8 $\pm$10$^{\circ}$ for the ellipticity and position angle respectively. Note that for an SIE, the projected mass density ellipticity $\varepsilon_{\rm M} \approx 3 \times \varepsilon_\Phi $, if the potential ellipticity $\varepsilon_\Phi << 1$ \citep{Golse2002}. The width of these priors are determined from the rms scatter in the \citet{Koopmans2006} sample, however we broaden the ellipticity prior by a factor of 5 in view of the best-fit ellipticity derived by \citet{Eisenhardt1996}. Furthermore, this is a higher redshift lens ($z = 0.893$) than the $z \sim 0.1-0.3$ SLACS lenses, and therefore would be less virialised from a statistical viewpoint. This is to say that given two elliptical lens galaxies, one at $z\sim0.1$ and the other at $z \sim 0.9$, there is a higher probability that the former is virialised as a result of hierarchal structure formation \citep[e.g.][]{Springel2005} and the difference in mergers rates at these two epochs \citep[e.g.][]{LeFevre2000}. We assume Gaussian, spherically symmetric source profiles. The Gaussian profile assumption was later relaxed, however, other Sersic indices did not improve the residuals. Following previous authors \citep[e.g.][]{Koopmans2006,Kormann1994} we assume a singular isothermal ellipsoidal potential (i.e. $\Phi \propto r^{-2}$ where $\Phi$ is the potential and $r$ is the radius). Deviation from this assumed density profile is likely our largest source of systematic error which is quantified in \S\ref{sec:systematics}.  Although {\sc glamroc} is able to model potentials with Sersic profiles, our attempts to free the inner density slope (assuming the \hst\,F814W data) did not yield successful results. 

The lens fitting is first performed with one SIE lens in a coarse grid mode where the  reduced-$\chi^2$ (denoted $\chi^2_\nu$ for the remainder of the paper) is recorded for the full range of plausible lens and source parameters. These results are used solely to make a rough approximation of the macro-lensing model. They are combined with an approximate measurement of the arc curvature, as well as the \citet{Eisenhardt1996} modelling results, to derive an Einstein radius Gaussian prior of $\theta_{\rm E}$~=~0.85~$\pm$0.1~arcsec. The source-plane centroid and scale radius have uniform (flat) priors. 

With priors in place, the algorithm is run. The MCMC routine employs the Metropolis-Hastings algorithm which compares the posterior probability of the current model with a candidate model.
We employ a Gaussian proposal distribution which is tuned to have an acceptance rate of 20\% of all proposal steps that result in a lower posterior probability, which allows local minima to be avoided. The full routine is set to run for $5 \times 10^{5}$ iterations, where convergence occurs after $\sim 1 \times 10^5$ iterations (i.e. sub-chains of this length exhibit equivalent statistics). A typical `burn-in'\footnote{The burn-in period refers to the number of MCMC iterations discarded before the overall statistics of the chain are determined. This is done because the algorithm takes an initial period to sufficiently explore the posterior, find the parameter sub-space of interest and exhibit statistics that are independent of the initial parameter vector.} period of $\sim2 \times 10^4$ iterations is observed. 

The MCMC routine must be run in uncorrelated parameter space to achieve convergence within a reasonable time-scale ($<$~24~hrs on a dual core 2.8~GHz machine). We derive the parameter vector covariance matrix by performing a long ($10^6$ iteration) run in correlated space. We then perform a principal component analysis (PCA) to transform the parameter vector into uncorrelated space where all trial steps are made. We use all Eigenvectors and Eigenvalues in the PCA analysis since the added computational time is negligible compared to the ray-tracing calculations and disk read/write speed.

\subsection{Lens Models}

\begin{figure*}
\includegraphics[width=1\textwidth]{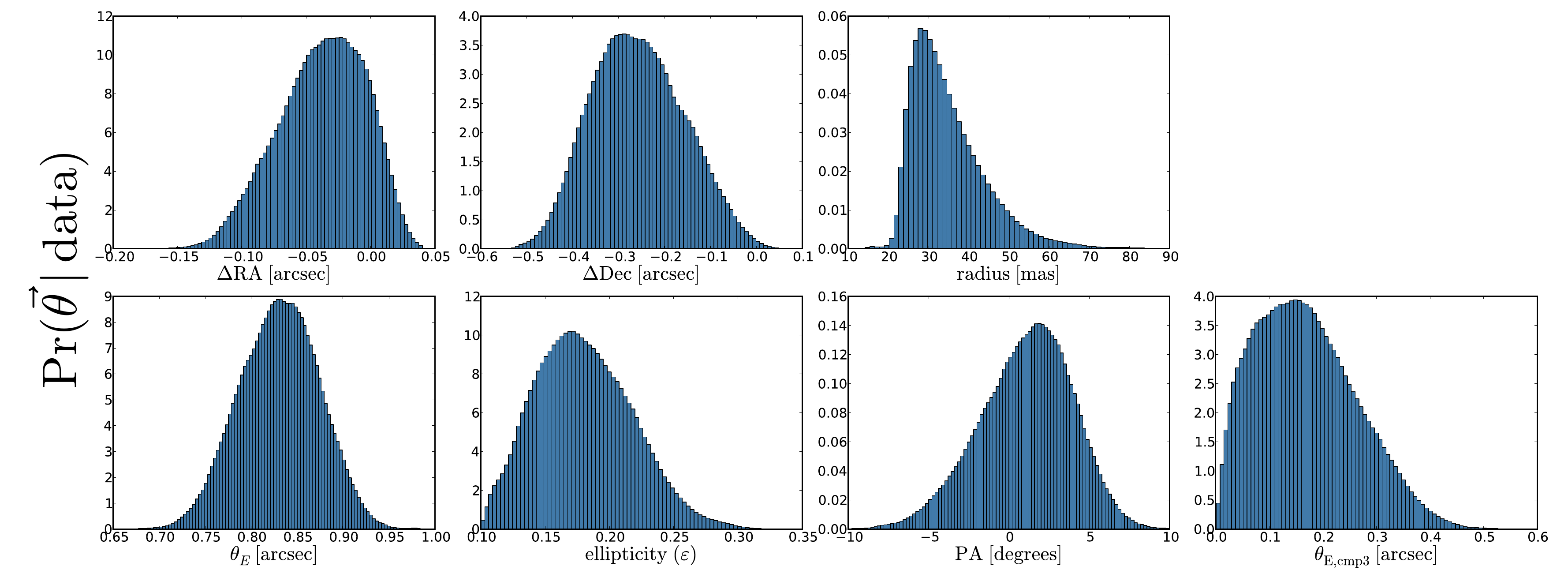} 
\caption{ The 7 histograms show the normalised, MCMC derived samples of the source (RA, Dec, scale radius) and lens parameters (Einstein radius, lens ellipticity, lens position angle, component 3 SIS Einstein radius), given the \hst\,F814W data. }
\label{fig:1Dhists}
\end{figure*}

The lens model derivation has followed three stages: 

\begin{enumerate}

\item  Lens Model 0: a single main lens (SIE); 
\item  Lens Model A: main lens (SIE) + an additional neighbouring mass (SIS) 
\item  Lens Model B: main lens (SIE) with free position + SIS neighbour.  

\end{enumerate} 

These three stages are motivated by the counter-image positional offset, the achieved $\chi^2_{\nu}$ values, as well as a measurement of the centre of arc curvature from both \hst\,F160W and \hst\,F814W maps. All of these are described in more detail below. Each stage has two runs, the first is performed using the process of `annealing' where the Gaussian proposal distribution FWHM decreases as a function of iteration number. This allows the full parameter space to be explored efficiently and is also used to determine the covariance matrix. Once the minima are found, the chain is re-run in uncorrelated space without annealing, using Gaussian proposal distributions, the variance of which are proportional to the corresponding Eigenvalues. The second run sets the initial parameter values to the best-fit of the first run. 

\subsubsection{Lens Model 0: Single SIE Lens}

This model has a total of 6 parameters: the unlensed source position ($RA,Dec$), source scale radius ($r_{\rm s}$), Einstein radius ($\theta_{\rm E}$), lensing potential ellipticity ($\varepsilon$), and its position angle ($PA$). The results of Lens Model 0  are surprisingly good (${\chi^2}_\nu \sim 32$) given the simplicity of the lens and source models. A single Gaussian source component models the majority of the flux in the arc, however there is a 150 mas North-East offset between the data and model counter-images. An analysis of all the individual model frames from the MCMC chain shows that there is no place within the parameter space that yields co-spatial data and model counter-images while simultaneously maintaining a reasonably symmetric arc. We disregard this lens model, since it clearly fails to reproduce the counter-image position not only in the best-fit model, but in all models in the allowable parameter space. Shear is a parameter that is routinely added in lens modelling and often leads to a dramatic improvement to the $\chi_\nu^2$, particularly for point sources. Shear was added to the lens model (requiring two additional two parameters: $\gamma_1, \gamma_2$) and marginally decreased the distance between the model and data counter-image to $\sim120$~mas, however it distorts the extended main arc, resulting in a $\chi_\nu^2$ that is almost unchanged. The best-fit shear magnitude was $\log_{10}{(\vert \gamma \vert)} = -1.94 \pm 0.55$ with a position angle towards component 3 ($\theta_{\gamma} \sim 30^\circ \pm 25^{\circ}$), as would be expected. This suggests that the inclusion of shear is not optimal for the extended emission present in the \hst\,F814W map, unlike the case of point sources.

\begin{figure*}
\includegraphics[width=0.9\textwidth]{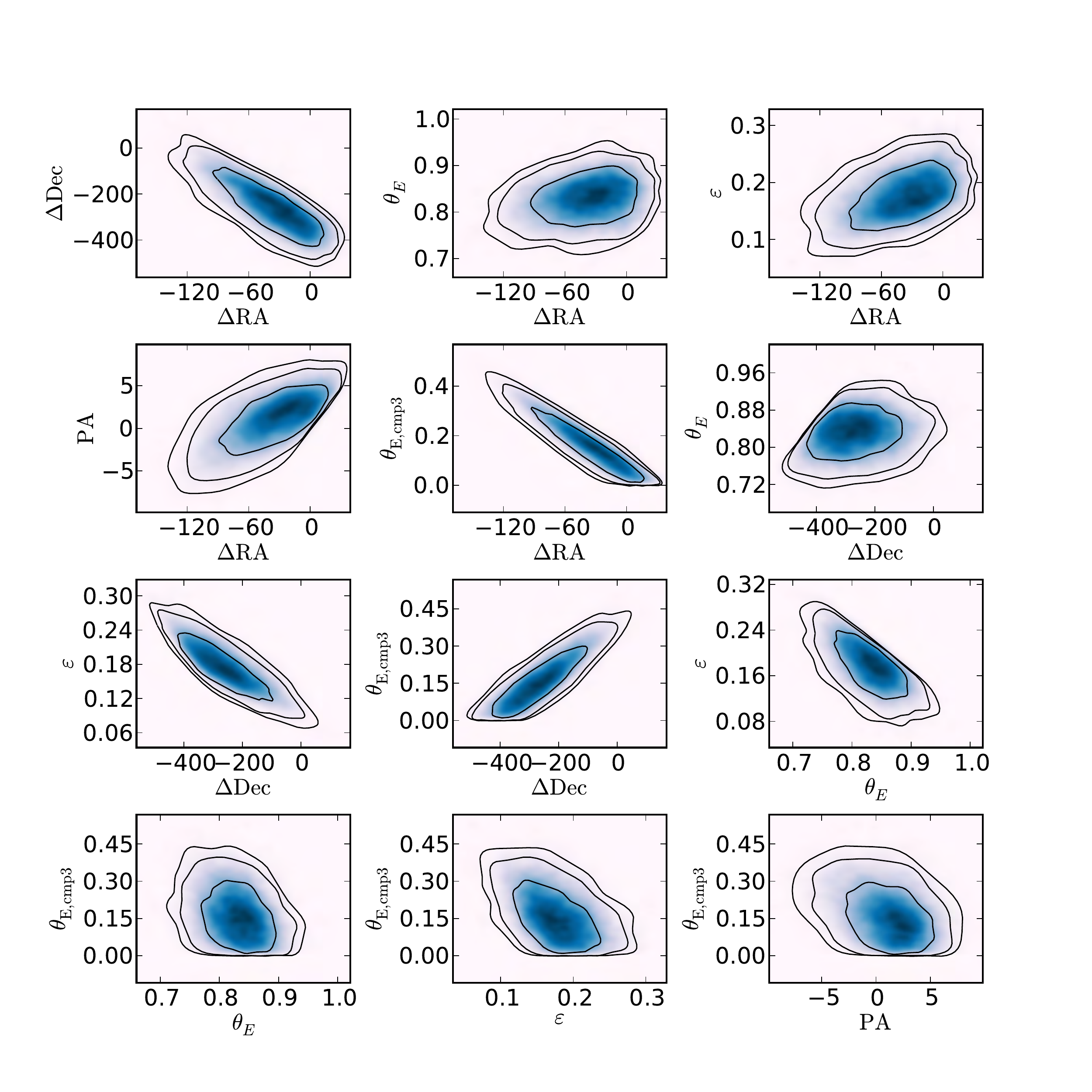} 
\caption{The 12 contours plots show a selected subset of the total of 21 2D marginalised PDFs of the source plane and lens parameters given the \hst\,F814W data. The selected two-dimensional posterior PDFs are those that showed any sign of degeneracy between the two parameters. The contours show the 68\%, 95\% and 99\% confidence levels. Note that plots are zoomed in to show the degeneracies clearer, in all cases the allowable parameter space is beyond the plot boundaries.  }
\label{fig:2Dhists}
\end{figure*}

\subsubsection{Lens Model A: SIE + satellite SIS}
To solve the counter-image position discrepancy encountered in Lens Model 0, we introduce an additional parameter in Lens Model A which comprises of the main lens SIE potential as well as a Singular Isothermal Sphere (SIS) potential to account for the influence for the nearby galaxy which is $\Delta \theta \sim$2.2~arcsec NNE from the primary lensing galaxy (component 3 in Fig.~\ref{fig:satgals}). A broad Gaussian prior PDF is adopted for the SIS velocity dispersion, based on the Faber-Jackson relation \citep{Faber1976}. The resultant Gaussian prior of the isothermal velocity dispersion has a mean $\sigma_v$ = 160~km\,s$^{-1}$ with a FWHM = 100 km\,s$^{-1}$ at a photometrically determined redshift of $z$ = 0.782 (Simpson~et~al., in preparation). Note that the mean of this prior is likely an over-estimate since closer inspection of the galaxy shows an slightly inclined morphology more representative of an S0 galaxy. A Faber-Jackson estimate of the stellar velocity dispersion based on the {\sl K-}band integrated flux ratio of component 2 and 3 (and the derived component 2 isothermal velocity dispersion) would result in an over-estimate of the component 3 mass since S0 galaxies have systematically lower $M/L_{B}$ ratios than ellipticals \citep[e.g.][]{Faber1979}. In addition, the photometrically derived redshift of component 3 ($z=0.782$) is essentially the most favourable lensing configuration for a source at redshift $z=2.3$ (due to their relative angular diameter distances). We therefore expect the mean of the component 3 isothermal velocity dispersion prior to be exaggerated, however a suitably broad prior alleviates this concern. 

This results in a model with 7 free parameters with physically motivated priors. The MCMC algorithm is tuned as described previously and run for $5 \times 10^{5}$ iterations. The one-dimensional marginalised posterior PDFs of all parameters are reasonably approximated by Gaussian distributions, which are plotted in Fig.~\ref{fig:1Dhists} and listed in Table~\ref{tab:lensmodel}. The two-dimensional posterior PDFs of all parameter pairs that show some degeneracy are also shown in Fig.~\ref{fig:2Dhists}. While the average $\chi^2_{\nu} = 30$ values are roughly the same for Lens Model A compared to Lens Model 0, the major difference is that the counter-image position has a far superior fit. There is still a $\sim 20-30$ mas southern offset between data and the model generated with the mean of all parameter posterior PDFs. However, it is clear that the additional SIS parameter results in putting the model counter-images in regions consistent with the data counter-image, justifying the additional model complexity. Further complexity could be added in the form of an external shear parameter, however we do not feel this is entirely necessary for our primary objectives of modelling dominant emission regions and their relative magnification, source plane position and size. 

The model makes two improvements on the \citet{Eisenhardt1996} lens model (apart from estimating uncertainties). Firstly, we use a more physical source model profile (Gaussian) than the uniform, constant surface brightness circular disk they employed. Secondly, we free the component 3 Einstein radius which they fixed to $\theta_{\rm E,cmp3} = 0.6$~arcsec, derived from the {\sl K}-band flux ratio of component 2 and 3 (which they assumed to both be at a redshift of $z=0.9$).  The derived Einstein radius of the component 3 galaxy greatly affects the component 2 Einstein radius and ellipticity, particularly since the redshift is not known. As a result, fixing the component 3 Einstein radius effectively rules out parts of the main lens parameter space. Our lens model has therefore explored a more extensive parameter space in its derivation.

\subsubsection{Lens Model B: SIE (free centroid) + satellite SIS}

\begin{figure}
\includegraphics[width=0.5\textwidth]{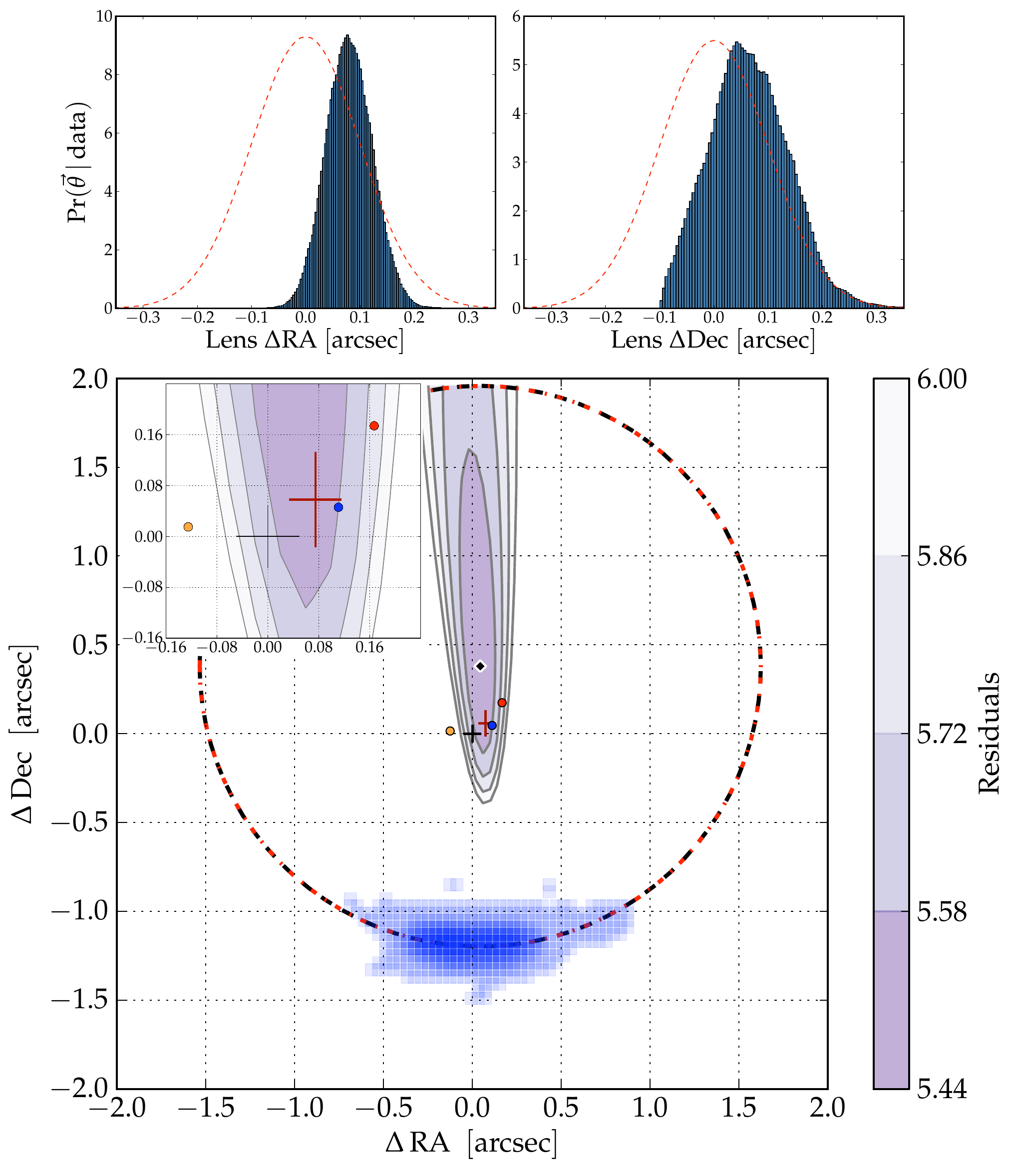} 
\caption{ {\bf Top:} Posterior PDFs of the lens potential centroid from Lens Model B. Over-plotted in red dashed lines are the assumed priors. {\bf Bottom:} \hst\,F814W curvature fit using both least squares (red dot-dashed circle) and orthogonal distance regression (black dashed circle) algorithms with the centre shown with a black diamond. The colour-scale shows the resulting $\chi^2_\nu$ residuals, which are generated by minimising the sum of the square of the distance (units: pixels) between each pixel and a circle. The inset is a zoomed in view of the \hst\,F160W Sersic fit centroid (black cross); the Lens Model B best-fit lens potential centroid (red cross); the \citet{Eisenhardt1996} \hst\,F814W curvature fit (blue dot); the \merlin 1.7~GHz lens emission peak (red dot); and the distinct core seen in the residuals of the \hst\,F160W Sersic fit (orange dot). Note the red cross represents the MCMC-derived uncertainty, however the black cross is purely illustrative, with a size $>>10$ times the formal Sersic fitting uncertainty.}
\label{fig:arcfit}
\end{figure}

This model is identical to Lens Model A (1 SIE, 1 SIS) however the main lens (SIE) potential centroid is allowed to vary, whereas before it was fixed to the \hst\,F160W map component 2 Sersic fit centroid. This model therefore has a total of 9 parameters. Varying the potential's centroid is primarily motivated by the \citet{Eisenhardt1996} comment that `The center of curvature of the arc was fitted and found to be 0.12'' west-northwest of the center of component 2'. The two additional parameters ($\Delta \, Lens \, RA$ and $Dec$) are assigned narrow Gaussian priors ($\pi(\theta) = 0.0 \pm 0.1$~arcsec). The algorithm is then run in the precisely the same manner as before, yielding some evidence for an offset between the stellar and lensing potential centroids. The derived values are listed in Table~\ref{tab:lensmodel}, and the fitted RA and Dec offsets are plotted in Fig.~\ref{fig:arcfit}. This plot shows that the enforced prior strongly influences the $\Delta \, Lens\,Dec$ posterior PDF, however if the prior is broadened, the MCMC algorithm does not converge. For the sake of brevity, the term `centroid offset' is referred to as the offset between any given component and the fixed lens potential centroid from Lens Model A. Again, the latter is derived from a Sersic fit to the \hst\,F160W map of component 2 and has co-ordinates (0,0) in all maps presented in this work. 

Lens Model B yields an average $\chi^2_{\nu} = 23$ that is 30\% lower than achieved by Lens Model A. Furthermore, it results in a component 2 lens potential ellipticity and position angle more closely aligned with the stellar distribution. As we will see in \S\ref{sec:cmp3}, the resultant component 3 mass and mass-to-light ratios are arguably more probable. The MCMC-derived centroid offset is broadly consistent with the curvature fitted value reported in \citet{Eisenhardt1996}. We repeat the curvature fitting here using two techniques to check consistency: a least-squares approach and an orthogonal distance regression algorithm. Both approaches yield the same results. The data are the pixel co-ordinates and the distance minimisation algorithm assigns the statistical weighting to different co-ordinate pairs by their normalised flux density. The fit is almost entirely insensitive to S/N clips and weighting schemes and both the \hst\,F160W and \hst\,F814W maps yield consistent fits, however the centre (black diamond in Fig.\ref{fig:arcfit}) is not consistent with the \citet{Eisenhardt1996} fit (blue dot). 

The lowest $\chi^2_\nu$ colour-scale tier in Fig.~\ref{fig:arcfit} clearly demonstrates that the adopted lens potential centroid (0,0) is not consistent with the residuals minima, however the MCMC-derived centroid (red cross) and \citet{Eisenhardt1996} estimate (blue dot) are. There are two more intriguing aspects of the data that are relevant. Firstly, there is some low level \merlin 1.7~GHz emission consistent with lens emission (as discussed in \S\ref{sec:1.6ghz}). The peak of this `lens emission' is $\sim 0.2$~arcsec north-west of fixed lens centroid and indicated by a red dot. Secondly, Sersic modelling of the \hst\,F160W lens emission reveals a distinct core that is offset from the Sersic component centroid. This component is assumed to be unresolved and so is modelled with a 0.1~arcsec PSF profile and has a fitted centroid $\sim 0.125$~arcsec east of the fixed potential centroid and is indicated on the map with an orange dot.

Based on the results above, we select Lens Model A for the remainder of this paper. While some evidence exists for offset baryon and lensing potential centroids, it does not appear convincing enough to add the complexity to the model. We therefore continue with Lens Model A, however will later derive the source magnifications and source plane parameters of both models to demonstrate the systematic uncertainties of this assumption. We set the lens parameters (component 2 Einstein radius, lens ellipticity and lens position angle; component 3 Einstein radius) to the mean of these 1D distributions for all subsequent modelling in this work. These fixed values are listed in Table~\ref{tab:lensmodel}. 

Finally, a lens modelling routine with a more flexible, pixelated source model \citep[as performed in e.g.][]{Warren2003,Vegetti2009,Suyu2009} was attempted but is currently computationally limited and is the topic of future work following optimisation of the source code.

\begin{table*}
\centering
\addtolength{\textfloatsep}{50mm}
\setlength{\extrarowheight}{4pt}
\begin{tabular}{c c  c  c  c c c c }
\hline
{\bf Parameter} & {\bf Prior} &  {\bf Lens Model A} & {\bf 68\% CL}   &  {\bf Lens Model B} & {\bf 68\% CL}   &{\bf Measured$^\dagger$} & {\bf Eis96} \\ 
\hline 
\hline
$x$                    & -- &   -0.025''     & $\pm$ 0.037''  &  -0.186''  & $\pm$ 0.035''  & -- & n/a \\
$y$                    &  -- &  -0.269''     & $\pm$ 0.11''  &  -0.0857''  & $\pm$ 0.095''  &-- & n/a \\
$r_{\rm s}$             &  --     &  0.028''  & $\pm^{0.009''}_{0.006''}$  &  0.031''  & $\pm^{0.012''}_{0.009''}$  &  -- & 0.010'' \\
$\theta_{\mathrm E}$      &  0.85$\pm$0.1''   &     0.827''    &   $\pm$0.044''     &       0.839''  & $\pm$ 0.040''  & --     &   0.82'' \\
$\varepsilon$            & 0.07$\pm$0.15      &       0.174    &   $\pm 0.042$ & 0.133  & $\pm$0.036  &     0.2   &   0.12   \\
$PA$                      &   -4.8$^{\circ} \pm 10^{\circ}$  &  1.9$^{\circ}$    &   $\pm 3.0^{\circ}$    &  -4.1$^{\circ}$  & $\pm 3.7^{\circ}$  &     -4.8$^{\circ}$    &   -11$^{\circ}$  \\
$\theta_{\mathrm{E,cmp3}}$ &   0.4$\pm$0.2''    &    0.124        &   $\pm$ 0.114''          &   0.379''  & $\pm$ 0.088''  &   --      &  0.6'' (fixed) \\
Lens $\Delta$RA          & 0.0$\pm$0.1''                 &           --        &   --            &     0.0749''    &   $\pm$0.042''  & -- & --  \\
Lens $\Delta$Dec          &  0.0$\pm$0.1''                &           --        &   --            &     0.058''    &   $\pm$0.074''  & -- & -- \\ 
  
\hline
\end{tabular}
   \caption{Lens Model A and B parameters derived from the \hst\,F814W map and their 68\% confidence levels as determined from their sampled posterior distributions. $x,y$ are the offsets between the source plane position and the {\sl observed} lens centroid; $r_s$ is the source plane scale radius; $\sigma_{\mathrm v}$ and $\sigma_{\mathrm v,\rm cmp3}$ are the 1D isothermal velocity dispersions for the main lens (component 2) and the line-of-sight galaxy (component 3); $\varepsilon$ is the intrinsic potential ellipticity of the main lens; $PA$ is the main lens position angle; $Lens \, \Delta RA,Dec$ is the lens potential offset from the \hst\,F160W Sersic fit centroid. The systematic uncertainty is discussed in \S\ref{sec:systematics}. The {\bf Prior} column lists the mean of each assigned Gaussian prior and the associated standard deviation. The {\bf Measured} column is based on GALFIT Sersic fits to the \hst\,F160W map. The final column ({\bf Eis96}) indicates the derived model values from \citet{Eisenhardt1996}. Note that the source plane parameters in \citet{Eisenhardt1996} were not published. We set the lens parameters to the mean of the 1D distributions for all subsequent modelling.\newline
${ }^\dagger$ This is the observed mass distribution ellipticity, as derived from a Sersic fit to component 2 in the \hst\,F160W map.}
   \label{tab:lensmodel}
\end{table*}

\subsection{Source Plane Modelling of IRAS~10214}

The same lens modelling procedure is now performed on the two resolved radio maps, and once again for the \hst\,F814W map. However, the lens model is now fixed to the Lens Model A values in Table~\ref{tab:lensmodel}, given the S/N of the radio data. We note that noise in the radio interferometric data includes correlated components, but we ignore this due to its low level and the dramatic gain in computational efficiency achieved by not performing the $\chi^2_{\nu}$ calculations in Fourier space\footnote{We note that a more sophisticated treatment of lens modelling, given interferometric data, allows a direct comparison with the measured visibilities and has the advantage of uncorrelated noise \citep[see][]{Wucknitz2004}.}. In holding the lens model constant, there are three free parameters in all fits: the source's centroid coordinates ($RA$, $Dec$) and scale radius $r_{\rm s}$. Uniform priors are assumed for these three parameters. Once again we assume Gaussian, spherically symmetric source profiles.

Our primary aim here is to place constraints on the relative source-plane properties of the three maps shown in Fig.~\ref{fig:maps}, using as few free parameters as possible. It is for this reason and the low S/N of the radio data that we assume that each component is modelled by a simple, spherically symmetric Gaussian profile. Our aim is not the perfect reconstruction of the source plane pixels, \emph{but rather the relative source plane configuration and sizes of the dominant components of the three different emission sources proposed}. 

\begin{figure}
\includegraphics[width=0.47\textwidth]{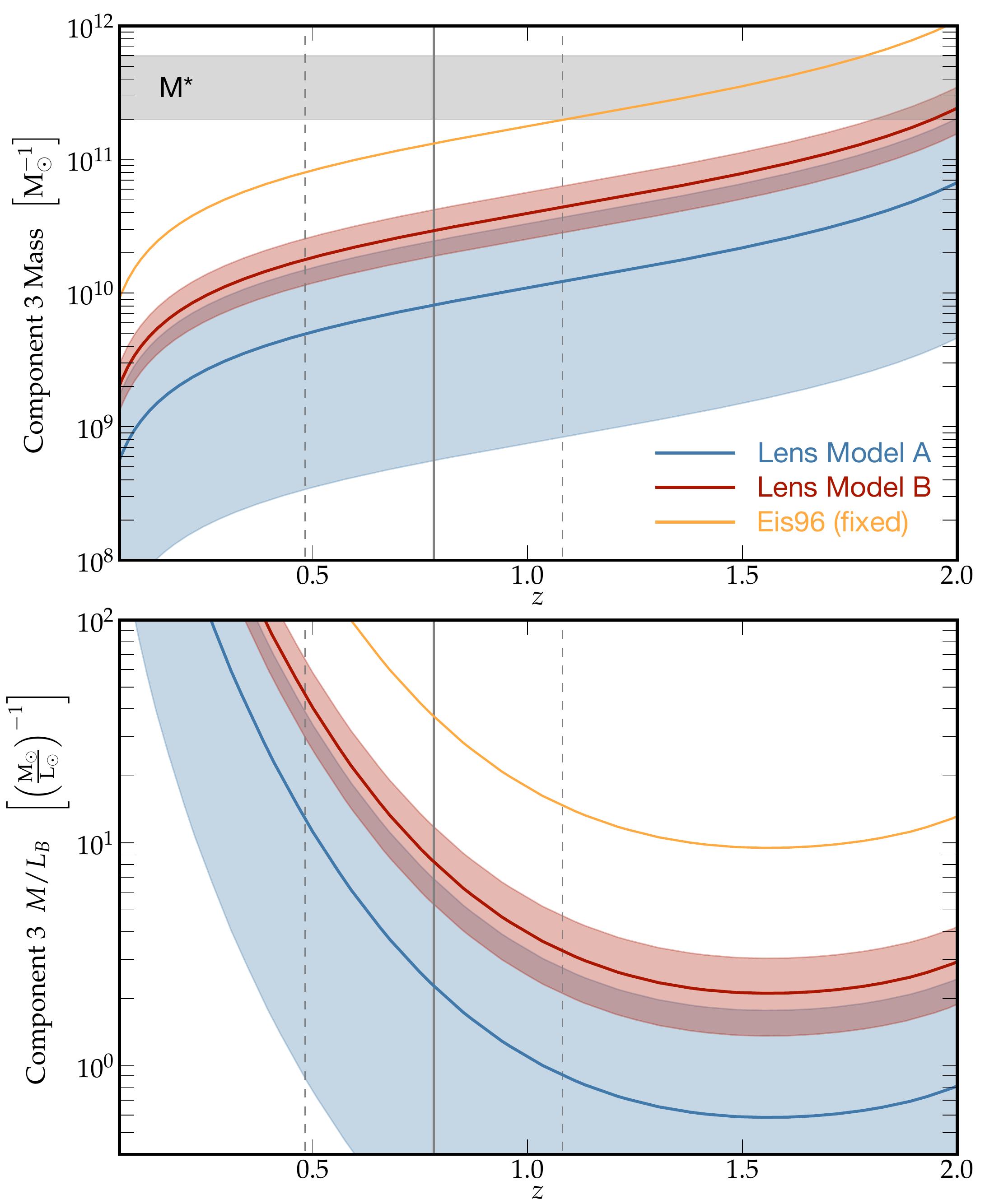}
\caption{ {\bf Top panel:} Component 3 mass (with Einstein radius) as a function of redshift, based on the Einstein radii derived in Lens Models A and B (blue and red respectively) shown with their MCMC-derived uncertainties in lower opacity. The orange line represents the fixed Einstein radius assumed in \citet{Eisenhardt1996}. The solid grey vertical line indicates the Simpson~et~al.~(in preparation) photometric redshift for component 3, with the rough 68\% confidence levels indicated by the dashed grey lines. The grey horizontal bar indicates an $M^*$ galaxy with an assumed mass-to-light ratio of 10-30. The angular diameter distances required are generated with the cosmological parameters listed in \S\ref{section_introduction}. {\bf Bottom panel:} Component 3 mass-to-light ($M/L_B$) ratio as function of redshift. Colour coding as in the top panel.  }
\label{fig:cmp3}
\end{figure}

\subsection{Component 3 Physical Properties}\label{sec:cmp3}

In Fig.~\ref{fig:cmp3} we show  the physical properties of component 3 based on our derived lens models. As stated before, the redshift is photometrically determined yielding large uncertainty in the intrinsic properties. Figure~\ref{fig:cmp3} (top panel) shows the mass of component 3 (inside the Einstein ring radius) as function of redshift. As can be seen, all three estimates are below $M^{*}$ ($\sim$10$^{11.3}$ M$_{\odot}$) for the most probable redshift range. The adopted Lens Model A has the lowest mass prediction which is $M_{\rm cmp3} \sim 8 \times 10^{9} \, {\rm M}_{\odot} \ \pm 0.5$~dex inside the Einstein radius of $\theta_{\rm E,cmp3} = 0.12$~arcsec. A consistency check is the derived mass-to-light ratio plotted in Fig.~\ref{fig:cmp3} (bottom panel). Rest-frame {\sl B}-band magnitudes  for the $z = 0-2$ range were interpolated from a high order polynomial fit to the F555W, F814W, F110W, F145M and F205W \hst filters. No stellar model can be assumed without a spectroscopic redshift, however the $L_B$ uncertainty is dominated by the lensing model and the not the assumed stellar spectrum. The three models all have plausible $M/L_B$ ratios (2--40) at the photometric redshift. Since this is in the dwarf galaxy regime, direct comparisons with large statistical samples are difficult, however the Lens Model B derived $M/L_B$ appears in the middle of the typical 3-30 range. Although Fig.~\ref{fig:cmp3} does not rule any of the three Einstein radii out, it does demonstrate that they are all physically plausible values at the photometric redshift.

\begin{figure*}[b]
\includegraphics[width=1\textwidth]{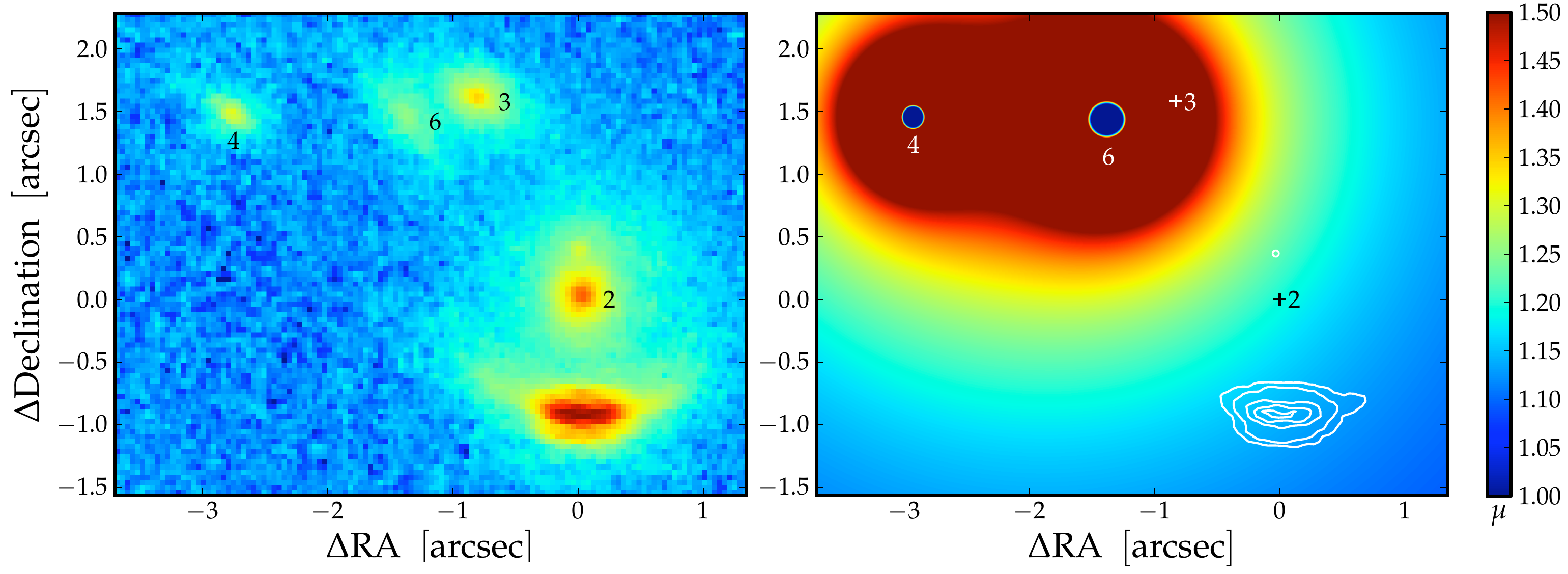}
\caption{{\bf Left:} Larger field of view of the IRAS~10214 region (\hst\,F160W map) showing three satellite galaxies to the north-east.
{\bf Right:} A map of the estimated magnification contribution from the component 4 and 6 galaxies, which are not included in our modelling. Note that the colour-scale is clipped between $1 \, < \, \Delta \mu \, < \, 1.5$, i.e. 0--50\% additional magnification. We argue this is an exaggerated estimate of the added magnification (see \S~\ref{sec:systematics} for a full description). The inner blue circle represents the saturated reverse parity magnification within each SIS critical curve. The numbered crosses indicate the centroids of the galaxies included in the lens model. }
\label{fig:satgals}
\end{figure*}

\subsection{Systematic Uncertainties}\label{sec:systematics}

The source of systematic error that is likely to dominate the error budget is the assumption of an isothermal density profile, which implies a fixed inner density power law slope $m$ = 1, where $m$ is defined in terms of the Einstein radius following \citet{Marshall2007} who show the convergence to be expressed by,

\begin{equation}
\kappa  =  \frac{2-m}{2} \left( \frac{\theta_{\rm E}}{\theta} \right)^m,
\end{equation}

\noindent in a generalised power law form. To quantify to the systematic error associated with this assumption we follow \citet[][Appendix A]{Marshall2007}. They quantify the inferred source size, $\Omega$, that results from small perturbations away from the isothermal logarithmic density slope ($m$ = 1). Their detailed appendix relates the inferred source size uncertainty ($\sigma_{\Omega}$) to the spread in the logarithmic density slope $\sigma_m$. To quantify the latter, these authors use the intrinsic spread of the power law indices fitted in the \citet{Koopmans2006} analysis of 15 early-type galaxies from the SLACS survey. We use the updated work in \citet{Koopmans2009}, which uses a larger sample of 58 early-type galaxies and finds an intrinsic spread of $\sigma_m$ = 0.2. With $\sigma_m$ quantified and the \citet{Marshall2007} relation between $\sigma_m$ and $\sigma_{\Omega}$

\begin{equation}
\frac{\sigma_{\Omega}}{\Omega}    = \mu \, \frac{\partial \mu^{-1}}{\partial m} \, \Big |_{m=1} \,  \sigma_m \, , 
\end{equation}

\noindent we find a systematic magnification uncertainty ${\sigma_{\Omega}}/{\Omega} \approx 2\sigma_m$ and so $\sigma_{\mu} \sim 0.4$. This 40\% uncertainty on the total magnification due to the density slope uncertainty is expected to dominate the systematic error budget given a particular lens model.

In principle, one could test our isothermal assumption by comparing the thickness of the arc and the north-south dimension of the counter-image. In the isothermal case, the two measurements have equivalent dimensions. However, in practice we could not get meaningful constraints on the counter-image thickness. We attempted this with the \hst\,F814W map, however there is significant uncertainty since the counter-image is located within the lens galaxy emission. Using GALFIT, the only case in which the Sersic fitting converges is if  the counter-image dimensions are fixed during the minimisation. For all results in this paper, these dimensions were fixed to that of the \hst\,F814W PSF. Attempts to marginally increase this to the north-south dimensions of the \hst\,F814W arc yields indistinguishable (quantitative and qualitative) residuals. This fitting challenge is compounded in the \hst\,F160W map since the host contamination is even greater. 

We have quantified the systematic uncertainty that results from the isothermal assumption, but it is also useful to consider the effect on source plane configuration. From the range of models considered in this work, it is clear that the Einstein radius and lens potential ellipticity share a degeneracy (as seen in their 2D posterior PDFs in Fig.~\ref{fig:2Dhists}). These two parameters counteract one another, in order to position the inner caustic so that the image-plane source positions are suitably reproduced. Variations to the inner density slope can be thought of in an analogous way: steeper inner-density slopes result in smaller Einstein radii which are compensated by larger ellipticity. Conversely, shallower density slopes equate to larger Einstein radii and therefore lower ellipticity. These statements are captured in the first order approximation in \citet{Marshall2007}, expressed as

\begin{equation}
\mu  \propto  \frac{1}{m^2 \, \epsilon}, 
\end{equation}

\noindent where $\epsilon$ is a small ($<< \theta_{\rm E}$) angular offset from the Einstein radius. In summary, changes to the inner density slope will cause shifts to the inner caustic position, which are compensated for by the lens ellipticity. The source-plane configuration is relatively similar in our three models. It is just the size of the source plane models (and hence the relative source-plane distances from one another) that change due to the changes in total convergence ($\kappa_{\rm tot}$). Departures from the isothermal profile will have similar results, the spread of which will reflect the observed spread in $m$. Stellar velocity dispersion measurements of the lens galaxy are required in order for this degeneracy to be broken, and for the inner density slope to be further constrained. This independent lens galaxy mass estimate will allow a more definitive mass model to be constrained, where the inner density slope is a free parameter in the lens modelling process.

Another source of systematic error is the neglect of two additional galaxies north-east of the main lensing galaxy. In Fig.~\ref{fig:satgals} (left panel) we show a larger field-of-view of the \hst\,F160W observation, revealing other galaxies in the field. Maintaining consistency with \citet{Eisenhardt1996} we label these galaxies component 4 and 6 (see Fig.~\ref{fig:satgals}; component 5 is the IRAS~10214 counter-image in their original paper). Component 4 has a spectroscopic redshift measured by the \citet{Goodrich1996} \mgii\ absorption detection at $z$ = 1.316, which is consistent with the photometric redshift of $z$ = 1.358 derived by Simpson et al.~(in preparation) using multiple \hst filters. Component 6 has neither a spectroscopic nor a photometric redshift. For the purpose of uncertainty estimation we assign it the redshift of its nearest neighbour (component 3 at $z$~=~~0.782). This is a conservative assumption, since this redshift has the highest lensing potential due to the source-lens redshift configuration (i.e. the relative angular diameter distances). The posterior PDFs of these photometric redshift estimates are non-Gaussian, however the 68\% confidence levels are of order $\Delta z$ = 0.3. The differing redshifts of the main lens and satellite galaxies are easily accounted for since the {\sc glamroc} software package is able to perform ray-tracing in multiple lens planes.

The Einstein radii of component 4 and 6 are estimated by scaling the main lens Einstein radius using the ratio of the {\sl K}-band magnitudes and the Faber-Jackson relation where mass $m \, \propto \sigma_v^4$ \citep{Faber1976} as well as accounting for their different redshifts\footnote{This assumes that the stellar velocity dispersion ($\sigma_v$) is equal to the modelled isothermal dark matter velocity dispersion.}. This over-estimates their Einstein radii since the primary lens and component 3 Einstein radii already include some convergence from component 4 and 6.

The additional magnification from component 4 and 6 is calculated by

\begin{equation}
\Delta \mu  =  \left[\{1 - (\kappa_{4} + \kappa_{6}) \}^2 - (\vec{\gamma}_{4}^2 + \vec{\gamma}_{6}^2)  \right]^{-1},
\end{equation}

\noindent where $\kappa_{4,6}$ and $\vec{\gamma}_{4,6}$ are the convergence and shear of component 4 and 6 respectively. In the right panel of Fig.~\ref{fig:satgals} we show the additional magnification from these two satellite galaxies. The colour-scale has been clipped between values of $\Delta \mu$ = 1.0 - 1.5 (i.e. $ \Delta \mu$ = 0 -- 50\%) for illustration. The figure shows that their combined potential results in a small additional magnification  at the 10--15\% level, bearing in mind that this is an exaggerated effect due to the reasons stated previously. This justifies our decision to neglect the potential of these galaxies and also illustrates why we were unable to get useful constraints on their velocity dispersions when an attempt was made to include them in the lens fitting procedure.

\section{Results}\label{sec:results}

\subsection{Source Plane Properties}

\begin{figure*}
\includegraphics[width=0.9\textwidth]{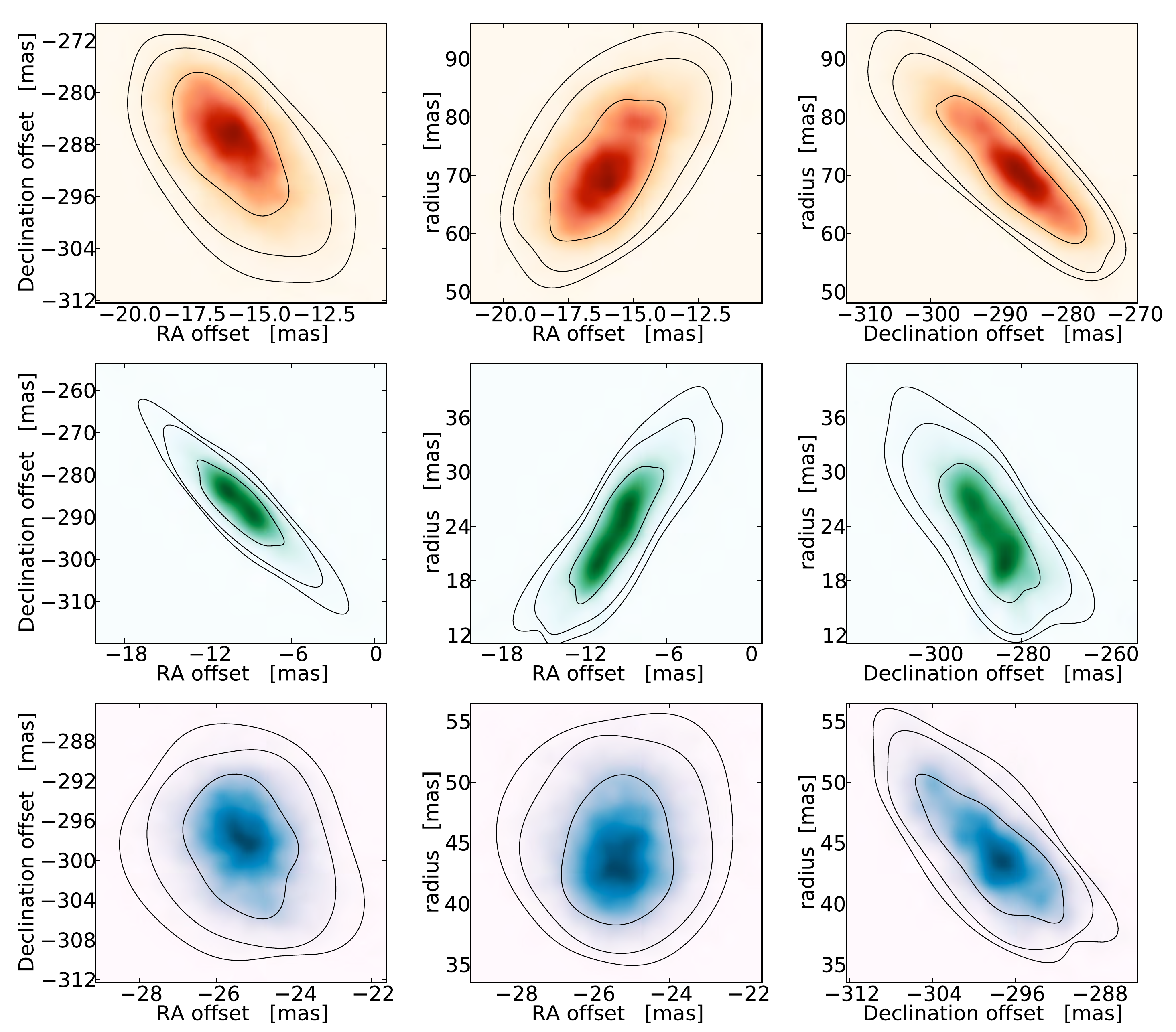} 
\caption{ The nine contour plots show the 2D marginalised posterior PDFs of the source plane parameters (RA, Dec, scale radius) of the system. The same colour coding is used for the three source components (1.7~GHz : red, 8~GHz : green, \hst\,F814W : blue). The contours show the 68\%, 95\% and 99\% confidence levels.}
\label{fig:likelihoods}
\end{figure*}

\begin{figure*}
\includegraphics[width=0.9\textwidth]{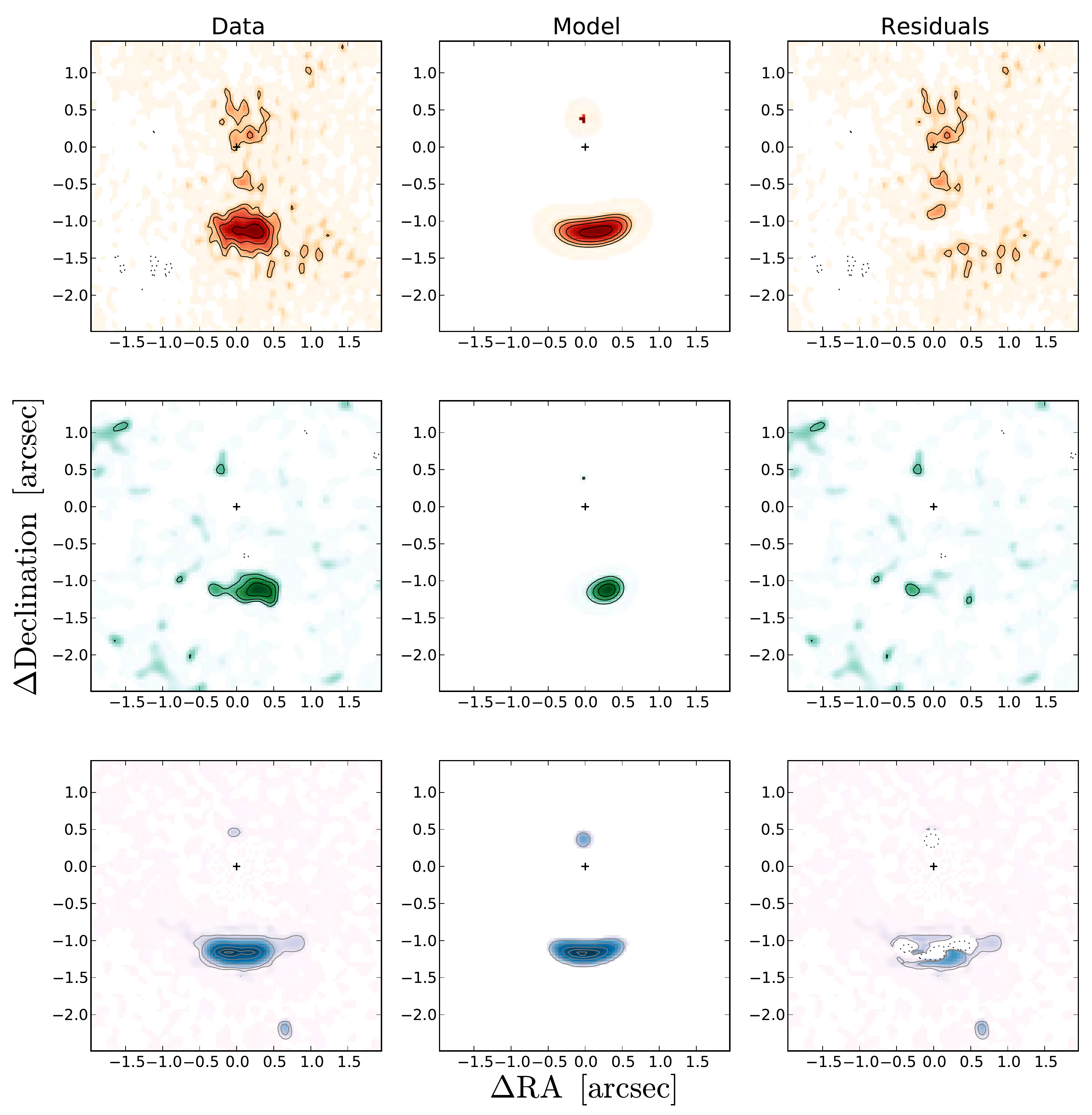}
\caption{ The nine images show the data, model and residuals of all three maps presented in Fig.~\ref{fig:maps}.  Note that the \emph{logarithmic colour scaling} is the same for all images in a row. The dominant component in the arc of the \hst\,F814W map is reasonably well modelled. The dashed lines indicate negative 3-$\sigma$ contours in the two radio maps and 10-$\sigma$ contours for the \hst\,F814W map. The crosses indicate the assumed lens potential centroid (\hst\,F160W Sersic centroid). The \hst\,F814W residuals suggest structure/clumpiness which is repeated in some of the trial MCMC models, however this could just as likely be intrinsic source structure and not the multiple images claimed by some deconvolutions of \hst maps. Over-plotted on the two radio maps are the {\sl unconvolved} model counter-images for illustration. These are multiplied by a factor 10. The 1.7~GHz residuals suggest some emission consistent with the lens itself, however there is also emission south-west of the arc which is not fit by the simple source model used here.} 
\label{fig:datamodelresidmaps}
\end{figure*}

 The source plane results that follow are entirely independent fits to each map, however the identical lens model is used in each case and each image plane is convolved with the appropriate PSF. Figure~\ref{fig:likelihoods} shows the 2D marginalised posterior PDFs of the source plane parameters. Each colour corresponds to one of the three observations (1.7~GHz : red, 8~GHz : green, \hst\,F814W : blue). We do not explore the {\sl AMI} 16 GHz and {\sl GMRT} 330 MHz source plane properties since they are both unresolved and the positional uncertainty is too large ($>>$ 0.1 arcsec) to make any meaningful constraints on the intrinsic position and radius. The resultant model and residual images for each of the 1.7~GHz, 8~GHz and \hst\,F814W image plane maps are shown in Fig.~\ref{fig:datamodelresidmaps}. 

The \hst\,F814W source plane model radius $r_{\rm s}$ = 44 mas (360~pc) is larger than that derived in \citet{Eisenhardt1996} ($2r_{\rm 814,Eis96}$ $\sim$ 11-20 mas), with a lower magnification of $\mu$ = 20$\pm 1$. However, direct comparison is difficult given that the Eisenhardt source model is a uniform surface brightness circular disk, that would lead to a smaller inferred radius. Furthermore, they assume an arc-to-counter-image flux ratio $\check{\mu}$ = 100, which appears to be their estimate of the magnification. We measure $\check{\mu}_{\rm 814,model} = 38$ which compares favourably with the derived `magnification' in previous work ($\check{\mu} \sim$ 50-100 in \citealt{Eisenhardt1996}) particularly since some differential extinction is expected at this wavelength since the counter-image light-path travels a factor $\sim$3 closer to the centre of the main lens galaxy, as argued in \citet{Nguyen1999}. To achieve the required arc-to-counter-image flux ratio, a $\Delta A_v = 0.23$ is required.

\begin{figure}[!h]
\includegraphics[width=0.47\textwidth]{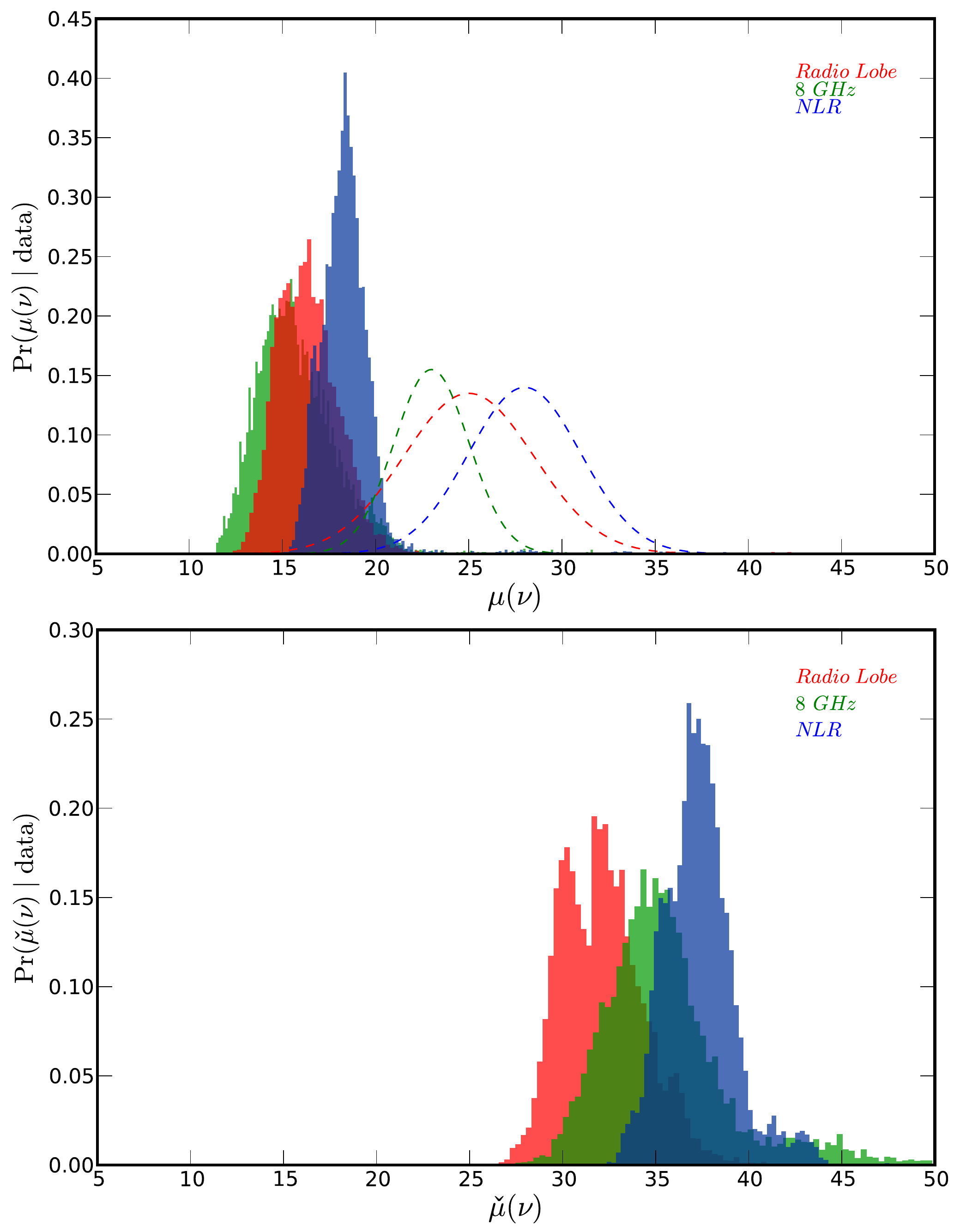} 
\caption{{\bf Top Panel:} Magnification posterior PDFs of the three maps based on the derived source plane reconstruction. The colours represent the same maps as before. The dashed lines indicate the same result for Lens Model B, which indicates the scale of the systematics ($\sim$40\%) in assuming co-spatial stellar and lensing potential centroids in this system. Additional systematics are derived in \S\ref{sec:systematics}. {\bf Bottom Panel:} Arc-to-counter-image flux ratio posterior PDFs of all three components. Note the significant offset when compared to the true magnification shown in the top panel. This demonstrates that $\check{\mu}$ does not accurately approximate magnification.  } 
\label{fig:munuHists}
\end{figure}

The radio lobe, which is presumed to dominate the 1.7~GHz map, is fit with an intrinsic scale radius $r_s$~=~72~mas~($\sim$600 pc) with a  magnification $\mu$ = 16$\pm1.5$ ($\check{\mu}_{\rm 1.7GHz,model}$ = 32), while the 8~GHz map, has the smallest source-plane scale radius ($r_s$ = 23 mas, 190 pc at z~=~2.3). Its size and position yield a magnification of $\mu_{8GHz}$ = 18$\pm2.5$. The rest-frame brightness temperature is calculated,

\begin{equation}
T_{\rm B}(\nu_{\rm e}) = \frac{c^2}{2 k_{\rm B}} \, \frac{S(\nu_{\rm o}) \, {\nu_{\rm e}}^{-2} \, (1 + z)^3}{\Omega}
\end{equation}

\noindent where $S(\nu_{\rm o})$ is the observed flux density, $\nu_{\rm e}$ is the rest frequency of the observed emission, and $\Omega$ is the apparent solid angle since surface brightness is conserved for lensed emission. The inferred brightness temperatures, $T_B \sim 1.1 \times 10^4$~K and $5 \times 10^3$~K for both radio lobe and 8~GHz components respectively, do not rule out star formation as the source of the radio emission. Typically, brightness temperatures need be above $T_{\rm B} \ >$ 10$^5$ K to exclude major contributions from thermal stellar emission and non-thermal supernova synchrotron emission. Note that the brightness temperature of a resolved radio component is conserved and therefore unaffected by lensing. In an analogous argument to that of brightness temperature we also calculate the source plane star formation rate density for comparison with the proposed value of a so-called `maximal starburst', $\Sigma_{\rm SFR}^{\rm max}$ $\sim$ 1000 M$_{\odot}\,$yr$^{-1}\,$kpc$^{-2}$ \citep[see ][]{Elmegreen1999,Thompson2005}. To calculate the implied star formation rates from measured radio luminosities, we follow \citet{Condon2002} who find the relation

\begin{equation}
SFR = 1.2 \times 10^{-21} \, L_{1.4GHz} \ \ {\rm M_\odot}\,{\rm yr}^{-1}
\end{equation}

\noindent  which is extrapolated to masses between $0.1{\rm M}_\odot < M < 100~{\rm M}_\odot$ assuming a Salpeter Initial Mass Function \citep[IMF,][]{Salpeter1955}. We calculate the luminosity at~1.4 GHz by applying the appropriate k-correction, and converting from higher frequencies by assuming a spectral index of $\alpha = 0.8$ (\begin{math} \alpha \equiv   - \log(S_1 / S_2) / \log( \nu_1 / \nu_2) \end{math}). If we assume that all radio emission is due to star formation, we derive lensed star formation rates of $SFR \sim 4 \times 10^4 \, \mu^{-1} \,  {\rm M}_{\odot}\,{\rm yr}^{-1}$ for both the 1.7~GHz and 8~GHz flux densities. Applying the derived magnifications and source-plane scale radii, we derive star formation rate densities of $\Sigma_{\rm SFR} = 2.0 \times 10^3$ and $2.1 \times 10^4 \ {\rm M}_{\odot}\,{\rm yr}^{-1} \, $kpc$^{-2}$, both above the theoretically motivated and observed saturation value of $\Sigma_{\rm SFR}$ which seems to suggest some radio core contribution. Note that $\Sigma_{\rm SFR}$ is unaffected by lensing, analogous to brightness temperature estimates. We use the source-plane radii as simpler estimates of the solid angle, and therefore use the unlensed star formation rates.

Inspection of the 8~GHz best fit model reveals an arc-to-counter-image flux ratio of $\check{\mu}_{\rm 8GHz,model}$ = 35, inconsistent with the $\check{\mu}_{\rm 8GHz,data}$ = 7  measured from the 8~GHz map, which is in-line with our assessment that the tentative counter-image is not a robust feature and therefore disregarded. Moreover, an analysis of the MCMC chain shows that the model counter-image is at no point within 100 mas of the 8~GHz peak. The nature of this component is perhaps the most puzzling, since if it is a radio core, we would expect it to be barely resolved. Three possibilities could lead to an exaggerated inferred size: convolution with a time and space varying atmospheric screen (smearing out the unresolved component); a significant flux contribution from star formation; as well as the possibility of small scale ($<$100 pc) jets. 

In Fig.~\ref{fig:munuHists} (top panel) we plot the magnification posterior PDFs of each source, given the fixed lens model. These three histograms show the confidence levels of preferential lensing in the system and that magnification $\mu$ is dependent on the emission region's size and proximity to the caustic. If these emission regions dominate different regions of the spectrum, it will appear as though magnification $\mu$ is a function of frequency $\nu$. The MCMC samples in Fig.~\ref{fig:munuHists} (top panel) were generated by taking the ratio of the image and source plane flux, where the source position and radius were taken from the original MCMC chain. The colours represent the same bands as before. We include the same magnification posterior PDFs that result from Lens Model B in dashed curves. At face value this suggests a systematic uncertainty of $\sim 40$\% , however this may be an over-estimate since the Lens Model B average $\chi^2_\nu$ is $\sim$30\% lower than that of Lens Model A. This makes a direct estimate of the systematic uncertainty that results from a fixed lens potential centroid more challenging. Nonetheless, this does demonstrate very similar \emph{relative} magnifications in two lens models, to go with the similar source plane configurations of both Lens Models A and B. These {\sl relative} values are the primary aim of this work and we argue that sufficient accuracy is achieved to accomplish this goal. In the bottom panel we plot the posterior PDFs of the arc-to-counter-image flux ratios for each source. The MCMC samples are drawn from each image plane model generated in the sampling of the magnification posterior PDFs (top panel) and so had higher spatial resolution and lower surface brightness thresholds set for higher accuracy. What the $\check{\mu}$ posterior plot demonstrates is that this is {\sl not} an accurate method to estimate magnification. 
 
In Fig.~\ref{fig:srcplane} we show the derived source plane model with contours of magnification over-plotted. The source plane reconstruction shows the scattered quasar light in a south-easterly direction to the 8~GHz flat-spectrum component. The steep-spectrum 1.7~GHz centroid is positioned roughly along the same vector between 8~GHz and \hst\,F814W centroids. The 1.7~GHz component has a larger size though, so the 8~GHz and \hst\,F814W components are essentially immersed within it. In Fig.~\ref{fig:srcplaneCLs} we show the same contours of magnifications, however in this case we plot the 68, 95 and 99 \% confidence levels of the centroid positions of the three components. This illustrates the degree of confidence we have on the perceived offsets between components, as well as the uncertainty of the macroscopic model, excluding systematic errors. This does not include the positional uncertainty as defined by $\sigma_\theta = 0.5\,$FWHM/(S/N), which is $\sim20$~mas for both radio maps. This positional uncertainty will have a larger effect on the magnification of the smaller 8~GHz component.

Note that the best fit model arc-to-counter-image flux ratios in all cases are significantly different from the true magnification derived by calculating the image-to-source plane flux ratio. This difference ($\sim$2-3) shows that the arc-to-counter-image ratio is an invalid approximation of the true magnification, despite its extensive use in previous work on IRAS~10214.

\subsection*{Susceptibility to Imprecise Astrometry}

\subsubsection*{Systematic Positional Uncertainty}

\begin{figure*}
\includegraphics[width=0.97\textwidth]{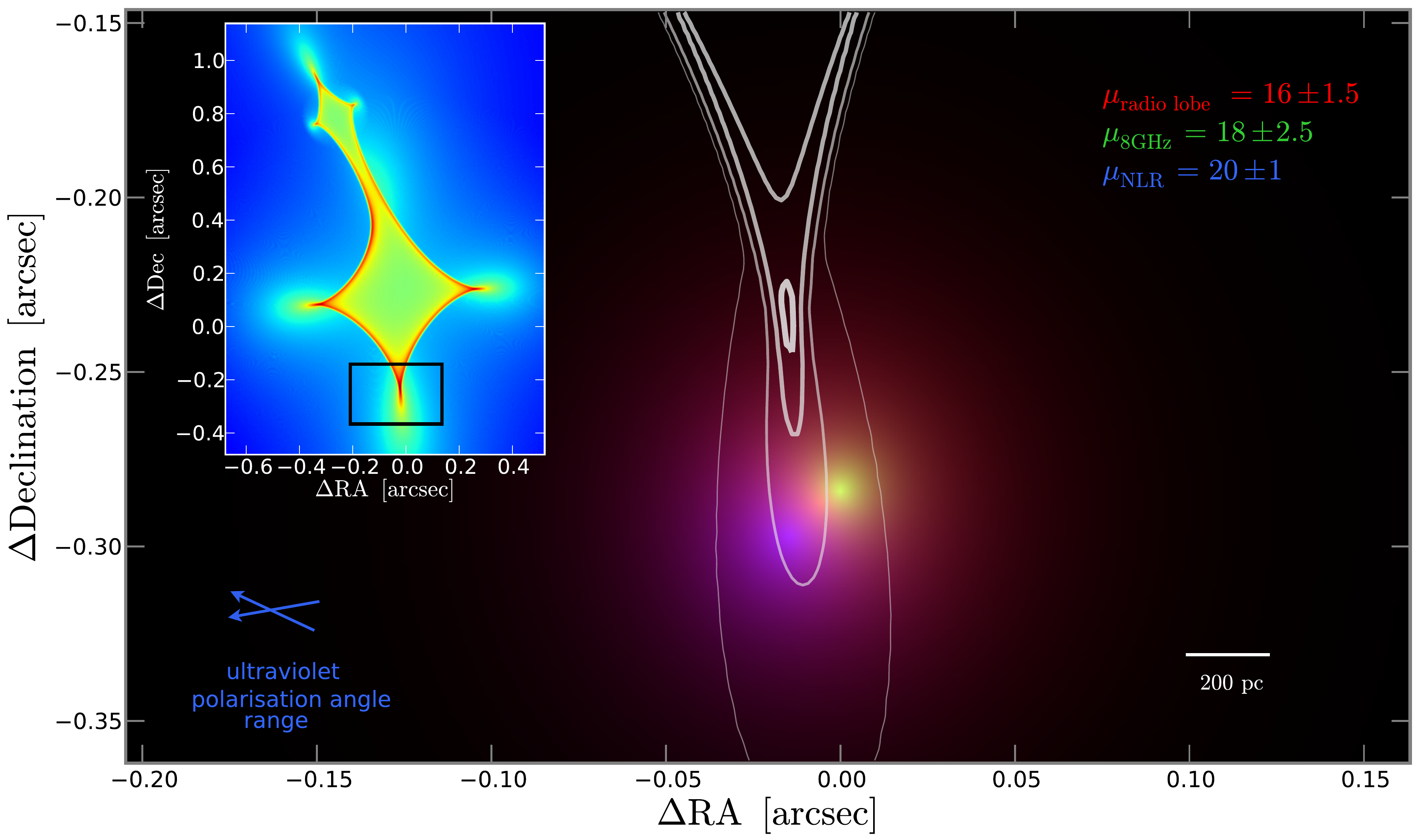} 
\caption{ Source plane reconstruction of IRAS~10214 showing the 8~GHz component (green), scattered quasar light (blue), and radio lobe (red). The white contours represent lines of equal magnification extending from the caustic at levels $\mu$ = 10, 20, 50, 100. The inset shows the full lens caustic with colour-scale representing magnification and the black rectangle showing the borders of the enlarged region. The magnification of each source could be computed from the convolution of each source with this magnification map. This is consistent with the method we use in practice, which is to integrate the model flux in the image plane and the source plane and take the ratio. }
\label{fig:srcplane}
\end{figure*}

\begin{figure*}
\centering
\includegraphics[width=0.97\textwidth]{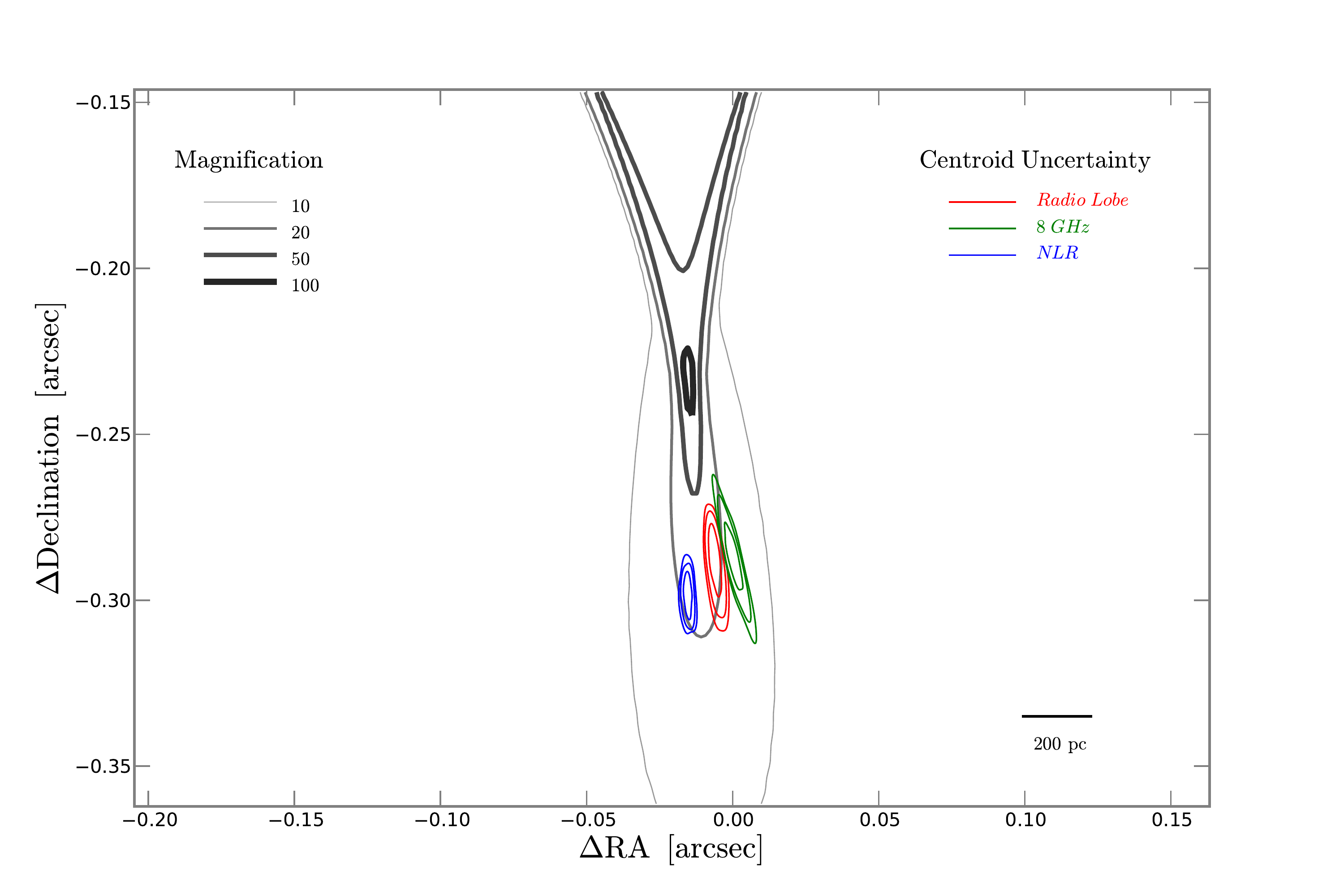} 
\caption{ Source plane with 68, 95 and 99\% confidence levels of the centroid position of each component shown in Fig.~\ref{fig:srcplane} (red, green, blue). Black contours indicate lines of equal magnification extending from the caustic at levels $\mu$ = 10, 20, 50, 100. }
\label{fig:srcplaneCLs}
\end{figure*}

Ockham's razor would suggest that the two radio components modelled here are most likely to be co-spatial, given the low S/N of the data. In \S2 we discuss the significant care we have put into demonstrating that astrometric accuracy is maximised and how \citet{Lawrence1993} find 1.4 and 8~GHz offsets consistent with ours. However, despite these two points, a natural question arises as to the sensitivity of the configuration in Fig.~\ref{fig:srcplane} to small shifts in astrometry. 

A relevant qualitative test is to determine if the two radio components could be co-spatial, given a reasonably large systematic offset in the appropriate direction. This is achieved by shifting the 8~GHz map south-east by 0.1~arcsec which is a relatively large astrometric error for 8~GHz \vla observations in A-array. The shifted map is then run in precisely the same MCMC routine as previously. The result is a similarly sized source plane radius ($r_s \sim 30$ mas) and source configuration, however with positional uncertainties that are consistent with the \hst\,F814W source plane component. Therefore, if two independent 8~GHz \vla observations separated by 4~years both suffered the same astrometric error then the 8~GHz component could be co-spatial with the \hst\,F814W component. 

An additional test was to enforce co-spatial source plane centroids for two or all three of the respective maps. Each trial image-plane map was generated, the $\chi^2_{\nu}$ calculated and then combined to determine the next MCMC step. This approach did not yield any meaningful results since no useful constraints could be made on a plausible source model with a common centroid (i.e. the data preferred no model at all since the output individual $\chi^2_\nu$ values worked antagonistically. 

We do not include positional uncertainties in our lens modelling. As calculated before, the \vla 8~GHz and \merlin 1.7~GHz position uncertainties are $\Delta \theta \lesssim 20$~mas. We constrain the added magnification uncertainty contribution from these random positional uncertainties by adding a 20~mas position dispersion in trial models prior to ray-tracing. This lowers the 8~GHz magnification by $\sim$10\%, and broadens the magnification uncertainty to 26\% ($\mu_{\rm 8GHz} = 16.1\pm4.7$), however the effect is significantly less on the 1.7~GHz component which retains its mean magnification with a minor broadening of the uncertainty ($\mu_{\rm lobe} = 16 \pm 2.0$). 

\section{Current View}\label{sec:currentview}

 Our source plane reconstruction (Fig.~\ref{fig:srcplane}) places strong constraints on the positions and radii of the three emission regions. These properties, and the inferred magnification, yield important information about the intrinsic nature of IRAS~10214. As discussed in \S1, the balance of the bolometric output has been the pertinent question in the case of IRAS~10214 since its discovery. Workers have invoked preferential lensing magnifications to explain some of the peculiar features of this galaxy. Through our lens modelling we find evidence of a slight preferential magnification of the NLR (and, by proxy, the AGN) when compared with the radio lobe. The scale of preferential lensing will be more accentuated when compared to larger scale emission regions like star formation. For example, a co-spatial 3 kpc Gaussian component undergoes a magnification $\mu$ = 7 (assuming the \hst\,F814W or 1.7~GHz centroids). This demonstrates a factor $\gtrsim$ 3 preferentially magnified AGN/NLR component.  

This preferential magnification of the AGN and NLR can qualitatively explain a range of mid-infrared features. Firstly, dust continuum models of IRAS~10214 have difficulty in reproducing the higher temperature dust ($T_{\rm dust} > 80$ K) implied by the flux levels in this spectral range ($\lambda_{\rm rest} < 40 \, \mu$m, \citealt{Efstathiou2006}). Secondly, the {\sl Spitzer} mid-infrared spectrum did not reveal PAH features which are expected to be strong given the substantial molecular gas mass ($M_{\rm H_2}\sim10^{11} \, \mu^{-1}  \, $M$_{\rm \odot}$), extremely large far-infrared luminosity, and the detection of heavy molecules/atoms (e.g. HCN (1-0), \citealt{VandenBout2004}; \ci, \citealt{Ao2008}). \citealt{Teplitz2006} suggest that preferential magnification of the AGN at a level of $\sim 3$ could suppress these PAH features, in line with the factor we determine here. It does not, however, explain the silicate emission feature revealed by this same {\sl Spitzer} spectrum. One possible scenario that would explain the silicate emission observed towards IRAS~10214 would require it's hot, silicate emitting clouds to preferentially lensed, but not obscuring the central AGN. Alternatively, special sight-lines can be invoked, that place a toroidal structure to be suitably inclined as to obscure the AGN but still exhibit silicate in emission \citep[see][]{Efstathiou2006}.

\section{Conclusions}

We have performed high resolution radio observations of a gravitationally lensed, starburst/AGN at $z = 2.3$. We have developed a Bayesian MCMC algorithm that performs forward ray-tracing on extended source models to constrain the lens model. Equipped with this model and the extensive multi-wavelength coverage, we investigate the source plane properties of our two resolved radio maps and the \hst rest-frame ultraviolet image and make the following conclusions. 

\begin{enumerate}

\item From the nature of this galaxy and its global spectral energy distribution we argue that the \merlin 1.7 GHz and \hst\,F814W maps are dominated by the radio lobe and scattered quasar emission respectively. The nature of the 8~GHz emission is likely a combination of supernovae and radio core/jet emission, which is explored in more detail in combination with VLBI observations presented in \citet{Deane2012c}. 

\item We find a 30\% preferentially lensed NLR over the radio lobe component owing to their different scales and proximity to the caustic. The NLR/AGN is a factor $\gtrsim$3 preferentially lensed over a co-located, 3 kpc radius emission component. This could explain some of the observed peculiarities in IRAS~10214 since it would mask the expected PAH features in the spectrum and would also contribute towards the measured mid-IR `excess' \citep{Efstathiou2006}.

\item This demonstration of preferential lensing shows that the brightest lenses may be the most affected since the survey selection band is likely to be the very wavelength that is preferentially magnified. This is true in the case in IRAS~10214 which was originally selected at rest-frame $\lambda_{\rm rest} \, \sim$ 18 and 30 $\mu$m, which is dominated by preferentially magnified, AGN-heated, hot dust. This will be an interesting point of comparison once similar studies are performed on the 500 $\mu$m selected {\sl Herschel} lenses \citep{Negrello2010}. 

\item We have demonstrated that the arc-to-counter-image flux density ratio is not an accurate estimate of magnification in this system, as has been used previously. It is incorrect by a factor $\gtrsim2$. 

\item There is some evidence for a gravitational potential centroid offset from the baryon centroid of the main lens (component 2). This increases the systematics substantially ($\sim40$\%), a source of uncertainty not usually considered in strong lensing analyses. This value may be an over-estimate however, since the $\chi^2_\nu$ value is 30\% lower in Lens Model B, so we are not comparing equivalent models. Despite the large systematics, the overall source plane configuration and global picture of a preferentially magnified NLR/AGN remains consistent.

\end{enumerate}

This paper has described the derivation of what we argue to be a more robust lens model for IRAS~10214. It has also described methods to quantify the relative magnification factors of different emission regions. What seems clear however, is that the radio maps presented here are not unambiguous tracers of physical components (i.e. star formation, jets, the active nucleus). As a result, we have undertaken observations that are clearer proxies of these components. We present spatially-resolved \co and 1.7~GHz VLBI imaging in \citet{Deane2012b,Deane2012c} respectively.

\section*{Acknowledgments}

We thank Ted Baltz for making his ray-tracing code ({\sc glamroc}) publicly available; Harry Teplitz for generously providing the calibrated {\sl Spitzer} IRS mid-infrared spectrum; Chris Simpson for allowing us to read a valuable pre-print. We thank Richard Saunders for setting up the {\sl AMI} observations; and Alejo  Mart\'\i nez-Sansigre, Aprajita Verma, Andreas Efstathiou and Natalie Christopher for valuable discussions. \merlin is a National Facility operated by the University of Manchester at Jodrell Bank Observatory on behalf of STFC. The National Radio Astronomy Observatory is a facility of the National Science Foundation operated under cooperative agreement by Associated Universities, Inc. We thank the staff of the {\sl GMRT} who have made these observations possible. The {\sl GMRT} is run by the National Centre for Radio Astrophysics of the Tata Institute of Fundamental Research. This research used the facilities of the Canadian Astronomy Data Centre operated by the National Research Council of Canada with the support of the Canadian Space Agency. This effort/activity was supported by the European Community Framework Programme 6 and 7, Square Kilometre Array Design Studies (SKADS), contract no. 011938; Advanced Radio Astronomy in Europe, grant agreement no.: 227290; and PrepSKA, grant agreement no.: 212243. PJM acknowledges support from the Royal Society in the form of a University Research Fellowship.


\label{lastpage}

\end{document}